\documentclass[useAMS,usenatbib,usegraphicx]{mn2e}

\def\gtsim {>\kern-1.2em\lower1.1ex\hbox{$\sim$}~}   
\def\ltsim {<\kern-1.2em\lower1.1ex\hbox{$\sim$}~}   
\newcommand{\iso}[1]{$^{#1}$}
\newcommand{\Msun}{M_\odot}

\def \apj {ApJ}
\def \apjs {ApJS}
\def \aj  {AJ}
\def \aap {A\&A} 
\def \mnras {MNRAS}
\def \pasj {PASJ}
\def \araa {ARA\&A}
\def \pasa {PASA}

\title[The Evolution of Isotope Ratios]{The Evolution of Isotope Ratios in the Milky Way Galaxy}
\author[Chiaki Kobayashi, Amanda I. Karakas, and Hideyuki Umeda]{Chiaki Kobayashi$^1$\thanks{E-mail: chiaki@mso.anu.edu.au}, Amanda I. Karakas$^1$, and Hideyuki Umeda$^2$\\
$^1$ Research School of Astronomy \& Astrophysics, The Australian National University, Cotter Rd., Weston ACT 2611, Australia\\
$^2$ Department of Astronomy, School of Science,
University of Tokyo, Bunkyo-ku, Tokyo 113-0033, Japan}

\begin{document}

\date{Accepted 2011 Feb 26. Received 2011 Feb 25; in original from 2010 Oct 20}

\pagerange{\pageref{firstpage}--\pageref{lastpage}} \pubyear{2010}

\maketitle

\label{firstpage}

\begin{abstract}
Isotope ratios have opened a new window into the study of the details of stellar evolution, 
supernovae, and galactic chemical evolution.
We present the evolution of the isotope ratios of elemental abundances (from C to Zn) in the 
solar neighbourhood, bulge, halo, and thick disk, using chemical evolution models with 
updated yields of Asymptotic Giant Branch (AGB) stars and core-collapse supernovae.
The evolutionary history of each element is different owing to the effects of the initial
progenitor mass and metallicity on element production. 
In the bulge and thick disk the star formation timescale is shorter than in the solar neighbourhood, leading
to higher [$\alpha$/Fe] ratios. Likewise, the smaller contribution from Type Ia supernovae in these
regions leads to lower [Mn/Fe] ratios.  Also in the bulge, the abundances of [(Na, Al, P, Cl, K, Sc, Cu, Zn)/Fe] 
are higher because of the effect of metallicity on element production from core-collapse supernovae.
According to our predictions, it is possible to find metal-rich stars ([Fe/H] $\gtsim -1$) that formed in the 
early Universe as a result of rapid star formation.
The chemical enrichment timescale of the halo is longer than in the solar neighbourhood, and consequently the
ratios of [(C, F)/Fe] and $^{12}$C/$^{13}$C are higher owing to a significant contribution from low-mass AGB stars.
While the [$\alpha$/Fe] and [Mn/Fe] ratios are the same as in the solar neighbourhood, the 
[(Na, Al, P, Cl, K, Sc, Cu, Zn)/Fe] ratios are predicted to be lower. Furthermore, we predict that 
isotope ratios such as $^{24}$Mg/$^{25,26}$Mg are larger because of the contribution from low-metallicity supernovae.
Using isotopic ratios it is possible to select stars that formed in a system with a low chemical 
enrichment efficiency such as the satellite galaxies that were accreted onto our own Milky Way Galaxy.
\end{abstract}

\begin{keywords}
Galaxy: abundances --- Galaxy: evolution --- stars: abundances --- stars: AGB and post-AGB --- stars: supernovae
\end{keywords}

\section{Introduction}
Elemental and isotopic abundances are the fossils of galactic archaeology.
Different elements are produced from stars on different timescales, therefore
elemental and isotopic abundance ratios provide independent information on the ``age'' of a system and 
can be used as a form of ``cosmic clock''. 
The formation and evolutionary history of galaxies can be constrained in theoretical models 
by using the information contained in the elemental abundances observed in stars
\citep[e.g.,][]{tin80,pagel1997,matteucci2001}.  
The space astrometry missions (e.g., GAIA) and large-scale
surveys (e.g., the high-resolution multi-object spectrograph HERMES on the Anglo-Australian
Telescope) will produce unprecedented information on the chemodynamical structure of the
Milky Way Galaxy. 
Theoretically \citet[hereafter K06]{kob06} succeeded in reproducing the average evolution of major elements 
(except for Ti) in the solar neighbourhood (see also the recent study by \citealt{rom10}), and 
\citet[hereafter KN11]{kob10} predicted the frequency distribution of elements in the Milky Way Galaxy 
depending on the location.

Isotope ratios of elemental abundances can provide more constraints not only on galaxy formation and evolution, 
but also on the detailed physics of Asymptotic Giant Branch (AGB) stars and supernovae.
The isotopic ratios of C, N, O, and Si have been measured in ancient meteorites and provide
information on conditions in the proto-solar nebula \citep[e.g., ][]{and93}, as well as the composition of
stellar winds from pre-solar grains \citep[e.g., ][]{zinner98}. Attempts have been made to use meteoritic
data to trace the chemical enrichment history of the Milky Way,
although this has been done only for a limited number of  elements
including O, Mg \citep{nittler08,nittler09}, and Si \citep{lugaro99,zinner06}.
The determination of isotopic ratios from stellar spectra requires very high quality data,
and isotopic determinations are only available for a small number of elements including 
Li, C \citep[e.g.,][]{spi06}, O \citep{smith90a}, Mg \citep{yon03,mel07}, Ti \citep{hug08}, 
Ba \citep{gallagher10}, and Eu \citep{sne02}. It should be noted however that from that list
only the carbon isotope ratio is relatively easy to obtain. For this reason there is data
for the carbon isotope ratio for stars in many different evolutionary states, metallicities,
and locations (including in external galaxies such as the Large and Small Magellanic Cloud).
For this reason, the $^{13}$C/$^{12}$C ratio is used to study the internal mixing and evolution 
of the observed stars since this ratio changes with time. These changes are predicted for 
low and intermediate-mass stars as well as for massive stars that evolve through the Wolf-Rayet phase.
With the next generation of telescopes, the study of isotopes will expand beyond the Milky Way and
neighbouring Magellanic Clouds toward the brightest stars in the outer halo or in neighbouring 
dwarf spheroidal galaxies. 
This will allow us to gain deeper insights into the galactic archaeology beyond 
our own local neighbourhood.

However, there are only a small number of theoretical predictions of the evolution of the isotopes
including the studies by \citet{tim95}, \citet{rom03}, \citet{fen03}, \citet{chi06}, and \citet{hug08}.
In this paper we present predictions for the time evolution of isotope ratios of elemental abundances (from C to Zn) 
in the Milky Way Galaxy. We do this by using the most up-to-date nucleosynthesis yields of core-collapse 
supernovae and AGB stars that are available.
In \S 2, we describe our chemical evolution models, and present new and updated supernovae yields. These new
yields are then compared to the AGB yields from \citet{kar10}.
In \S 3, we show the time/metallicity evolution of elemental abundances and isotope ratios for the solar 
neighbourhood, bulge, halo, and thick disk.
We focus on the average evolution of abundances in the Galaxy and assume that the abundances of the 
carbon-enhanced metal-poor stars \citep[CEMP, ][]{bee05} are explained with other effects 
such as inhomogeneous enrichment, faint supernovae, and binary effects.
We end with conclusions in \S 4.

\section{The Model}

\subsection{Chemical Enrichment Sources}

We include the latest chemical enrichment input into our chemical evolution models as follows.
The basic equations of galactic chemical evolution are described in \citet[hereafter K00]{kob00} and K06, 
where the instantaneous recycling approximation is not applied, i.e., the contributions from stars of all mass ranges 
are computed as a function of the initial masses and metallicities of stars.

{\bf Stellar winds} ---
The envelope mass and pre-existing heavy elements are returned by stellar winds from all dying stars.
From stars with initial masses of $M \ltsim 1M_\odot$, it is assumed that no new metals are ejected,
and that only the material outside the He core is returned to the interstellar medium (ISM), 
which contains elements with the abundance pattern of the Galaxy at the time when the stars formed.
The He core mass is set as $M_{\rm remnant}=0.459$ and $0.473M_\odot$ for $0.7$ and $0.9M_\odot$, respectively.

{\bf Asymptotic Giant Branch (AGB) stars} ---
Stars with initial masses between about $0.8-8 M_\odot$ (depending on
metallicity) pass through the thermally-pulsing AGB phase. The
He-burning shell is thermally unstable and can drive mixing between
the nuclearly processed core and envelope. This mixing is known as 
the third dredge-up (TDU), and is responsible for enriching
the surface in $^{12}$C and other products of He-burning,
as well as elements heavier than Fe produced by the {\em slow} neutron
capture process \citep{busso99,herwig05}.
Importantly, the TDU can result in the formation of a C-rich envelope,
where the C/O ratio in the surface layers exceeds unity.  In AGB stars
with initial masses $\gtsim 4\Msun$, the base of the convective 
envelope becomes hot enough to sustain proton-capture nucleosynthesis 
(hot bottom burning, HBB).  HBB can change the surface 
composition because the entire envelope is exposed to the hot 
burning region a few thousand times per interpulse period. 
The CNO cycles operate to convert the freshly synthesized  $^{12}$C
into {\em primary} $^{14}$N, and the NeNa and MgAl chains may also
operate to produce $^{23}$Na and Al.  AGB stars with HBB have short 
lifetimes ($\tau \ltsim 100$~Myr) and are one of the stellar sites 
proposed as the polluters of globular clusters 
\citep[e.g., ][]{cottrell81,renzini08}, even if quantitative
problems with the models exist \citep[e.g.,][]{fen04}.
Overall a large fraction of light elements such as C, N and F are
produced by AGB stars, while the contribution toward heavier elements 
(from Na to Fe) is negligible, except perhaps for specific isotopes 
(e.g., $^{22}$Ne, $^{25,26}$Mg), in the context of galactic
chemical evolution. AGB stars are also an important source of elements
heavier than Fe \citep{travaglio01,travaglio04}. 

The nucleosynthesis yields of \citet{kar10} were calculated from
detailed stellar models, where the structure was computed first and
the nucleosynthesis calculated afterward using a post-processing algorithm. 
Yields are included for 77 nuclei including all stable isotopes from H
to \iso{34}S, and for a small group of Fe-peak nuclei. The details of
this procedure and the codes used to compute the models have been
previously described in some detail, see for example \citet{karakas09}
and references therein. All models were evolved
from the zero-age main sequence to near the tip of the thermally
pulsing AGB. The TDU efficiency governs the nucleosynthesis in the 
lower mass models; this was found to vary as a function of H-exhausted
core mass, metallicity, and envelope mass \citep[see][for
details]{karakas02}.  For example,
in the $Z=0.02$ models, no TDU was found for $M \le 2\Msun$. 
For the intermediate-mass models, the TDU was found
to be efficient and the occurrence of HBB 
also played a strong role in determining the final yields.
The occurrence of HBB also strongly depends on the initial
mass and metallicity, with HBB occurring in lower mass stars 
with a decrease in metallicity (at 3$\Msun$ at $Z = 10^{-4}$
whereas it only starts at $\sim 5\Msun$ at $Z=0.02$). 
Furthermore, HBB is eventually shut off by the action of 
mass loss.

The main uncertainties affecting the nucleosynthesis yields of AGB stars 
involve convection and mass loss.  The models employ the mixing-length
theory of convection with $\alpha = 1.75$. On the first giant branch, 
Reimer's mass loss is adopted with $\eta_{\rm R} = 0.4$. On the AGB, 
\citet{vw93} mass loss is used for most models, with the exception of 
the intermediate-mass 3 to 6$\Msun$ with $Z = 10^{-4}$ models,
where we adopt Reimer's mass loss \citep[see][for details]{kar10}.

The main difference between the \citet{kar07} and \citet{kar10} yields
is the choice of reaction rates used in the post-processing
algorithm.  \citet{kar10} used an updated set of proton and
$\alpha$-capture rates that include some of the latest experimental 
results for important reactions involved in the CNO cycle, NeNa and MgAl
chains. Furthermore, \citet{kar10} assumed scaled-solar initial
abundances for the $Z=0.008$ and $Z = 0.004$ models. In contrast, \citet{kar07}
adopted initial abundances for the Large and Small Magellanic
Clouds from \citet{russell92} which are sub-solar for C, N,
and O. The updated reaction rates of the $^{22}{\rm Ne}(p,
\gamma)^{23}{\rm Na}$, $^{23}{\rm Na}(p, \gamma)^{24}{\rm Mg}$, and 
$^{23}{\rm Na}(p, \alpha)^{20}{\rm Ne}$ reactions result in 
$\sim 6$ to 30 times less Na is produced by intermediate-mass models 
with HBB. Note that with the updated yields the Na overproduction
problem found by \citet{fen04} may be solved (but not the O depletion
required stars in globular clusters, although see models by 
\citet{ventura09}).

We take the AGB yields for $M=1.0$, 1.25, 1.5, 1.75, 1.9, 
$\sim 2$\footnote{$2.1 M_\odot$ model is provided for $Z=0.004$ and 0.008.},
2.25, 2.5, 3.0, 3.5, 4.0, 4.5, 5.0, 5.5, 6.0, and $6.5
M_\odot$\footnote{$6.5 M_\odot$ model is available only for $Z=0.02$.}
and $Z=0.0001$, 0.004, 0.008, and 0.02 from \citet{kar10}.
We use the yields for $Z=0.02$ at $Z \ge 0.02$. The yields of the radioactive
isotopes $^{26}$Al and $^{60}$Fe are added to $^{26}$Mg and $^{60}$Ni
yields, respectively.
The masses of white dwarfs, $M_{\rm remnant}$, are also taken from
\citet{kar10}, and the material outside the white dwarfs ($M_{\rm
initial}-M_{\rm remnant}$) is returned to the ISM via stellar winds, which contains the
newly produced metals (processed metals) and the initial metals that
existed in the progenitor stars (unprocessed metals).

The AGB yields ($m_i$) are defined as the difference between the amount of the species 
in the wind and the initial amount in the progenitor star.
Therefore, the yields of some isotopes (e.g., $^{15}$N) are negative because they are 
destroyed during stellar evolution. However, in chemical evolution models, it is possible 
that the abundances of such elements at time $t$ is lower than the adopted initial abundances, 
which causes numerical problems. This is often the case for $^{15}$N, and we set the $^{15}$N 
yield to be 0 if it is negative.
In chemical evolution models, we define the mass of ejecta as the summation of processed 
metals $M_{\rm ejecta} \equiv \sum_{i=^2{\rm H}}^{^{62}{\rm Ni}} m_i$. 
The rest ($M_{\rm initial}-M_{\rm remnant}-M_{\rm ejecta}$) contains unprocessed metals, 
of which the abundance pattern is not scaled with the solar abundance, but is the abundance 
pattern of the galaxy at the time when the stars formed.

For $Z=0$, theoretical models of stars undergo violent evolutionary episodes not seen at higher metallicities. 
The ingestion of hydrogen leads to an H flash, followed by a ``normal'' He-shell burning phase.
We take the yields and remnant masses from \cite{cam08} for $M=0.85, 1.0, 2.0$, and $3.0 M_\odot$. 
For Na, we assume that the yield is reduced by a factor of 10 because the old reaction rates were 
adopted in the calculations. At $M > 3M_\odot$, no metals are produced.
This assumption may not be valid but does not affect the average chemical evolution of galaxies.

{\bf Super AGB stars} ---
The fate of stars with initial masses between about $8-10 M_\odot$ is uncertain.
The upper limit of AGB stars, $M_{\rm u,1}$, is defined as the minimum mass for carbon ignition, and is 
estimated to be larger at high metallicity and also at low metallicity than at $Z\sim10^{-4}$ \citep{gil07,sie07}.
At $M>M_{\rm u,1}$, stars may produce some heavy elements and may explode as so-called Type 1.5 
supernovae, although no such supernovae has yet been observed (K06).
Above this mass range, stars may explode as electron-capture supernovae \citep{nom84,kit06}, 
but the metal production (lighter than Fe) is predicted to be very small.
In our models, we set $M_{\rm u,1}=4,6.5,6.5,6.5$, and $7M_\odot$ for $Z=0$, 0.0001, 0.004, 0.008, and 0.02, 
respectively, and assume that no metals are produced from 
$M_{\rm u,1}$\footnote{The remnant mass at $M_{\rm u,1}$ is extrapolated for $Z<0.02$.} to $10M_\odot$.
The remnant mass $M_{\rm remnant}$ is set as $1.01, 1.12$, and $1.15M_\odot$ for $7, 8$, and $10 M_\odot$, respectively.

{\bf Core-collapse supernovae} ---
Although a few groups have presented feasible calculations of exploding $10-25 M_\odot$ stars \citep{mar09,bru09},
the explosion mechanism of core-collapse supernovae (Type II, Ib, and Ic Supernovae) is still uncertain.
However, the ejected explosion energy and $^{56}$Ni mass (which decays to $^{56}$Fe) can be directly estimated 
from the observations, i.e., from the light curve and spectral fitting of individual supernova.
As a result, it is found that many core-collapse supernovae ($M \ge 20M_\odot$) have an explosion
energy that is more than 10 times that of a regular supernova ($E_{51}\gtsim10$, \citealt{nom06}), as well as producing more
iron and $\alpha$ elements (O, Mg, Si, S, Ca, and Ti). These are called as hypernovae (HNe).
The fraction of HNe is uncertain and we set $\epsilon_{\rm HN}=0.5$ at $M\ge20M_\odot$.

\begin{figure}
\includegraphics[width=8.5cm]{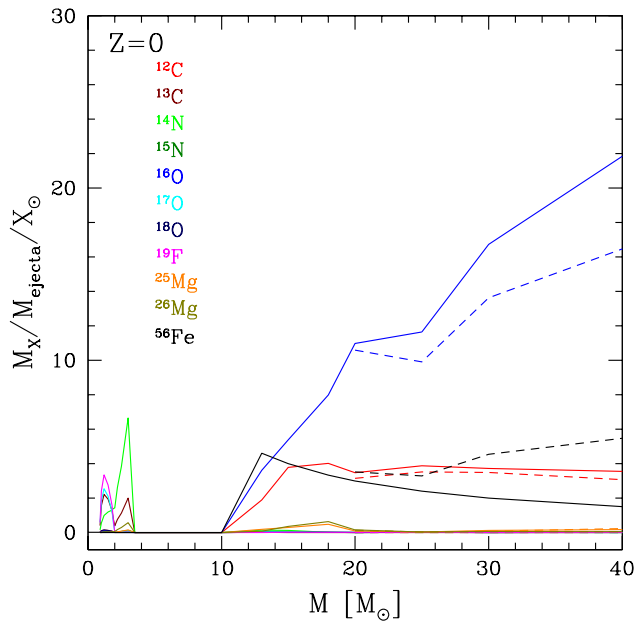}
\caption{\label{fig:yield1}
Mass fraction elements in the ejecta for $Z=0$ as a function of progenitor mass, 
normalized to the solar abundances \citep{and89}.
The solid and dashed lines are for SNe II and HNe, respectively.
}
\end{figure}

\begin{figure}
\includegraphics[width=8.5cm]{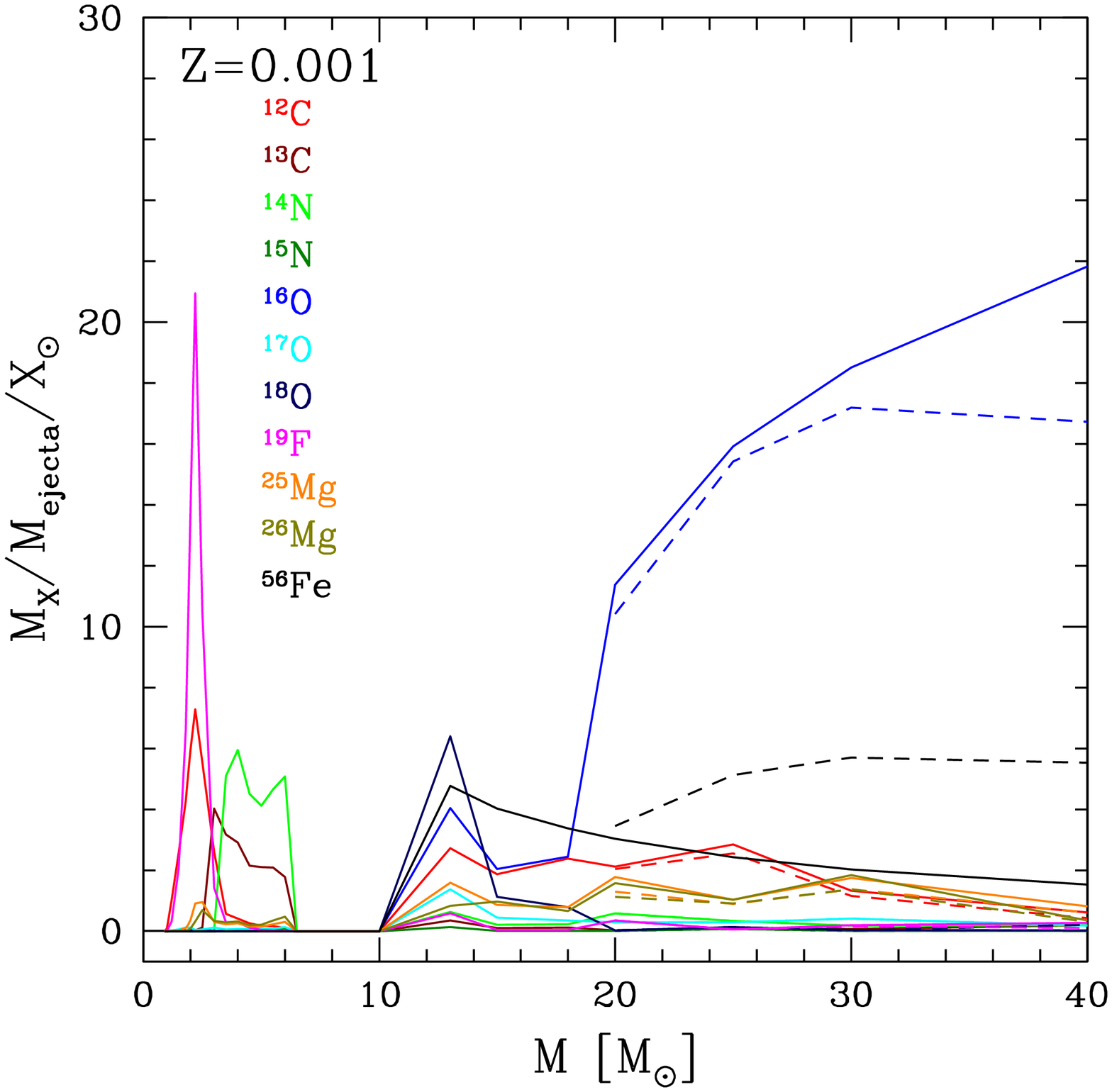}
\caption{\label{fig:yield2}
The same as Fig.\ref{fig:yield1} but for $Z=0.001$.
}
\end{figure}

\begin{figure}
\includegraphics[width=8.5cm]{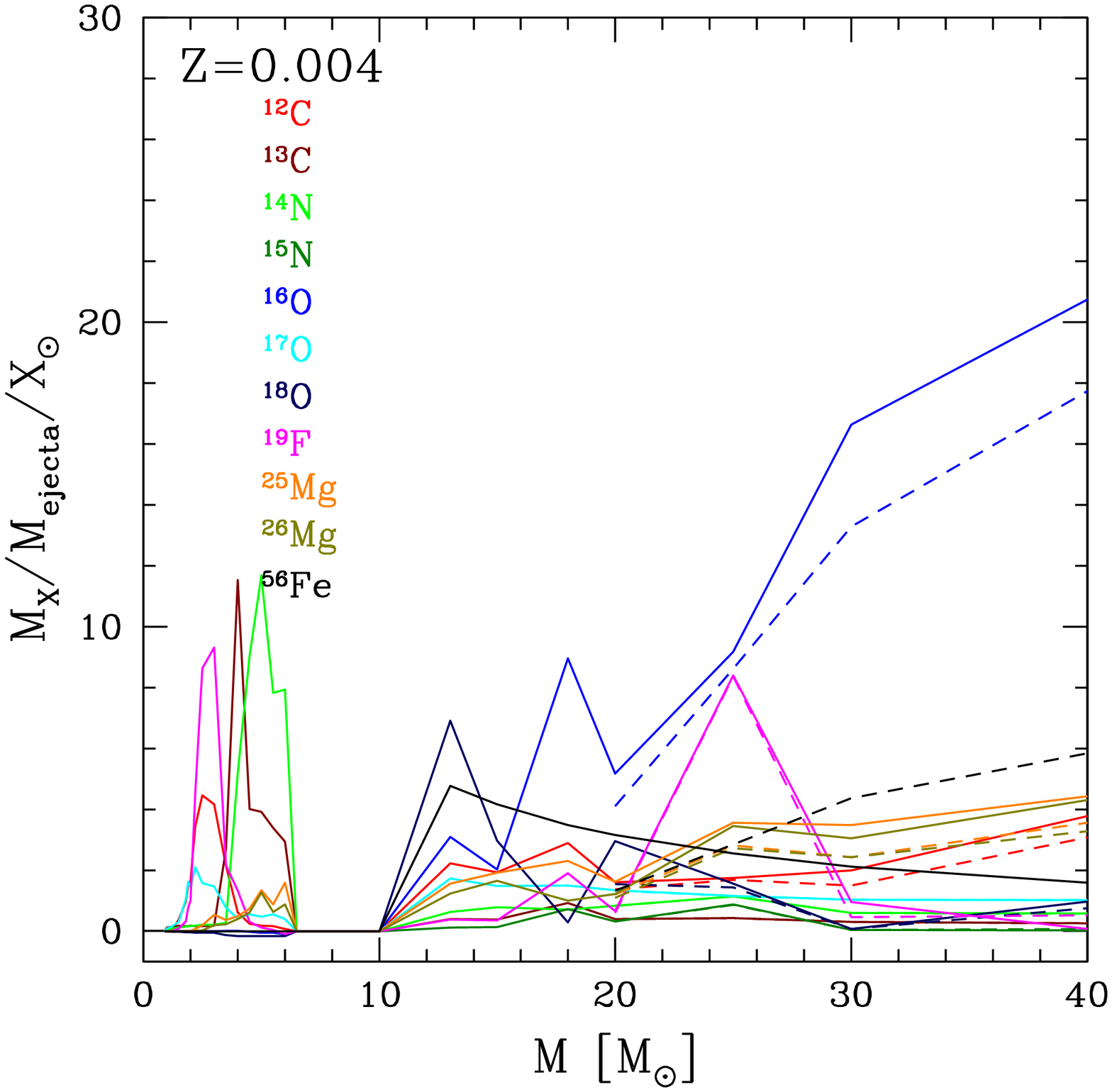}
\caption{\label{fig:yield3}
The same as Fig.\ref{fig:yield1} but for $Z=0.004$.
}
\end{figure}

\begin{figure}
\includegraphics[width=8.5cm]{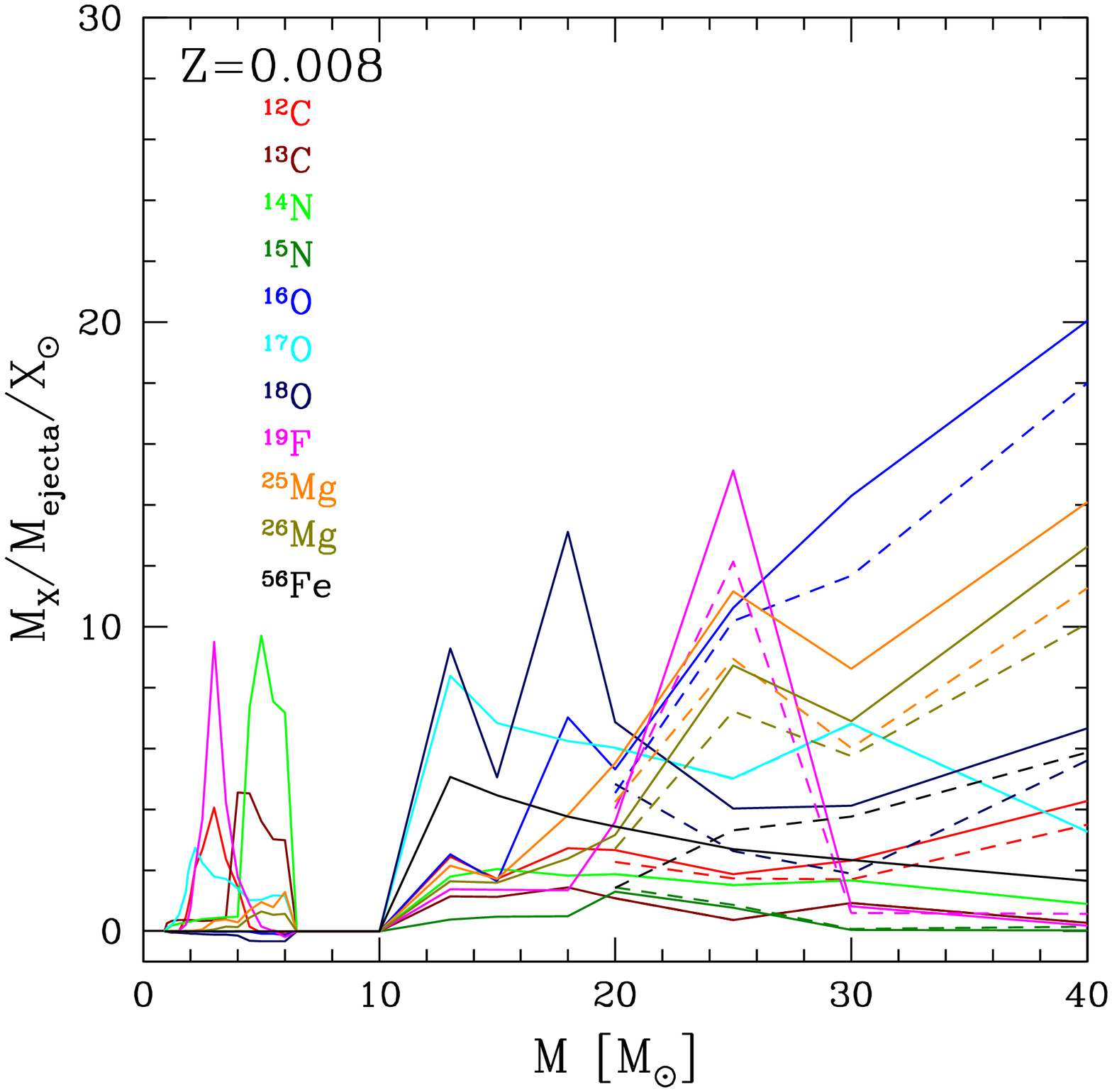}
\caption{\label{fig:yield4}
The same as Fig.\ref{fig:yield1} but for $Z=0.008$.
}
\end{figure}

\begin{figure}
\includegraphics[width=8.5cm]{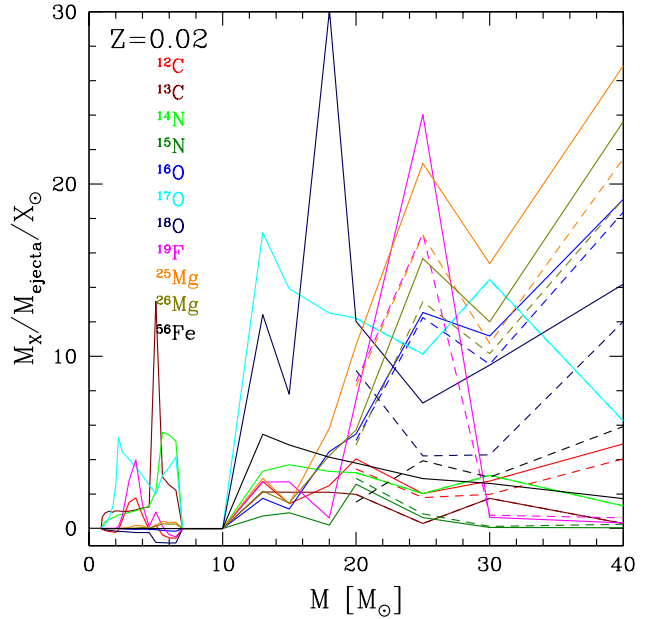}
\caption{\label{fig:yield5}
The same as Fig.\ref{fig:yield1} but for $Z=0.02$.
}
\end{figure}

\citet{kob06} presented the nucleosynthesis yields of SNe II and HNe as a function of the 
progenitor mass ($M= 13$, 15, 18, 20, 25, 30, and $40 M_\odot$) and metallicity ($Z= 0$, 0.001, 0.004, and $0.02$).
In terms of isotope ratios, three models, [$M=18M_\odot$, $Z=0.004$] and [$M=25M_\odot$, $Z=0.02$] for 
SNe II and HNe, showed a relatively large production of $^{13}$C and N. This was caused by the convective 
mixing of hydrogen into the He-burning layer. 
The cause of this mixing is uncertain and it does not always manifest itself in the stellar models.
For this reason, we re-calculated the progenitor star models and explosive nucleosynthesis. 
The results of these new calculations without such mixing are presented in Table \ref{tabsn}
and shown in Figs. \ref{fig:yield1}-\ref{fig:isotope5}.  While the updated yields of $^{13}$C and N are reduced compared to the results of
K06, the yields of other major isotopes are not significantly different.

As in K06, the yield tables provide the amount of processed metals ($m_i$) in the ejecta (in $M_\odot$).
The mass of the ejecta is given as 
$M_{\rm ejecta} \equiv \sum_{i=^1{\rm H}}^{^{74}{\rm Ge}} m_i=M_{\rm final}-M_{\rm cut}$. 
Stellar winds reduce the stellar mass to $M_{\rm final}$ at the onset of the supernova explosion, with
the central mass $M_{\rm cut}$ falling onto the remnant.
The stellar winds with the mass of $M_{\rm initial}-M_{\rm final}$ contains unprocessed metals, which are 
not included in the tables but are included in the chemical evolution models. 
The contribution from the stellar winds are added (Eq.8 in K00) to the supernova ejecta (Eq.9 in K00).
The abundance pattern of the stellar winds is not scaled with the solar abundance but is the abundance 
pattern of the galaxy at the time when the stars formed.
Note that newly synthesized $^{4}$He in the winds is not included and thus we will leave the evolution of helium to a future study.

The upper limit of core-collapse supernovae, $M_{\rm u,2}$, is not well known owing to uncertainties in 
the physics of blackhole formation.
We set $M_{\rm u,2}=50M_\odot$, which is constrained from the [$\alpha$/Fe] plateau at [Fe/H] $\ltsim -1$ (K06).
Because the envelope mass that contains $\alpha$ elements is larger for massive stars, a 
larger $M_{\rm u,2}$ results in higher [$\alpha$/Fe] ratios.
For $M>M_{\rm u,2}$, we assume that no metals are produced because the central part of the progenitor is 
likely to collapse to form a blackhole.

{\bf Rotating massive stars} ---
The rotation of stars induces the mixing of C into the H-burning shell, which produces a large amount of 
primary nitrogen \citep{mey02,hir07}.
Rotation also affects the CO core mass, and a few groups are calculating the stellar evolution of 
rotating stars from the main sequence through to the final explosive supernova stage, although the yields
are not yet available. The rotational velocity of these stars is typically a free parameter.
To show the effect of rotation, in one of our models we include the yields 
of $^{3,4}$He, $^{12,13}$C, $^{14}$N, $^{16,17,18}$O, and $^{22}$Ne
from \citet{hir07} for $Z=0$ with [$20M_\odot$, 600 km s$^-1$], [$40M_\odot$, 700 km s$^-1$], 
and [$85M_\odot$, 800 km s$^-1$], with $M_{\rm u,2}=100M_\odot$.

{\bf Pair-Instability supernovae} ---
Stars with $100M_\odot \ltsim M \ltsim 300M_\odot$ encounter the electron-positron pair instability 
and do not reach the temperature of iron photodisintegration.
Pair-instability supernovae (PISNe) are predicted to produce a large amount of metals such as S and Fe.
This contribution is not included in our chemical evolution models because the number of such stars are 
expected to be very small and because no signature of PISNe has been detected in metal-poor stars \citep{ume02,ume05}.
Recently, \cite{kob10b} showed that the observed elemental abundance pattern of very metal-poor Damped Lyman $\alpha$ systems is 
inconsistent with the yields of PISNe, and is instead consistent with the yields of faint core-collapse supernovae.

{\bf Type Ia Supernovae (SNe Ia)} ---
The progenitors of the majority of SNe Ia are most likely 
Chandrasekhar (Ch) mass white dwarfs (WDs).
For the evolution of accreting C$+$O WDs toward the Ch mass,
two scenarios have been proposed.
One is the double-degenerate scenario, i.e., the merging of double C$+$O WDs 
with a combined mass surpassing the Ch mass limit.
However, it has been theoretically suggested that this leads to accretion-induced collapse rather 
than SNe Ia \citep{nom91}, and the lifetimes are too short to reproduce the chemical evolution in 
the solar neighbourhood (\citealt{kob98}, hereafter K98; \citealt{kob09}, hereafter KN09).
The other is the single-degenerate (SD) scenario,
i.e., the WD mass grows by accretion of hydrogen-rich matter via mass transfer from
a binary companion.  
The mass accretion rate is limited to trigger carbon deflagration \citep{nom82}, but the allowed parameter 
space of binary systems can be significantly increased by the WD wind effect if the metallicity is 
higher than [Fe/H] $\sim -1$ (K98).

In our models, based on the SD scenario, the lifetime distribution function of SNe Ia is calculated 
with Eq.[2] in KN09, taking into account the metallicity dependence of the WD winds (K98) and 
the mass-stripping effect on the binary companion stars (KN09).
There are two kinds of progenitor systems. One is the main-sequence$+$WD system with timescales of 
$\sim 0.1-1$ Gyr, which are dominant in star-forming galaxies (the so-called prompt population). 
The other is the red-giants$+$WD system with lifetimes of $\sim 1-20$ Gyr, which are dominant in 
early-type galaxies.
Although the metallicity effect of SNe Ia has not yet been confirmed by supernova surveys, it is 
required to account for the presence of a young population of SNe Ia, which in turn are required
by chemical evolution of the Milky Way Galaxy (KN09).
Note that the observed elemental abundance pattern (e.g., the low [Mn/Fe] ratios) in dwarf spheroidal 
galaxies are more consistent with the enrichment from low-mass SNe II than that 
from the SN Ia enrichment (K06).

For SNe Ia we take the nucleosynthesis yields from \citet{nom97}.
The metallicity dependence of the progenitors is not included but is not expected to be very large 
(H. Umeda, K. Nomoto, et al., private communication).
Note that Ni is overproduced at [Fe/H] $\gtsim -1$
compared to the observations in the solar neighbourhood,
but this can be solved by tuning the propagation speed of the burning front and the 
central density of the white dwarf \citep{iwa99}.

\subsection{Yields}

\begin{figure}
\includegraphics[width=8.5cm]{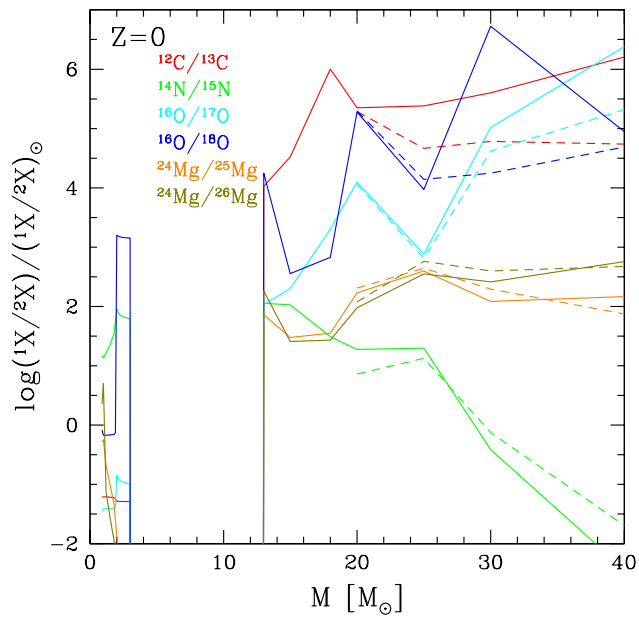}
\caption{\label{fig:isotope1}
Mass ratios of isotopes in the ejecta for $Z=0$ as a function of progenitor mass, 
normalized by the solar ratios \citep{and89} in the logarithmic scale.
The solid and dashed lines are for SNe II and HNe, respectively.
}
\end{figure}

\begin{figure}
\includegraphics[width=8.5cm]{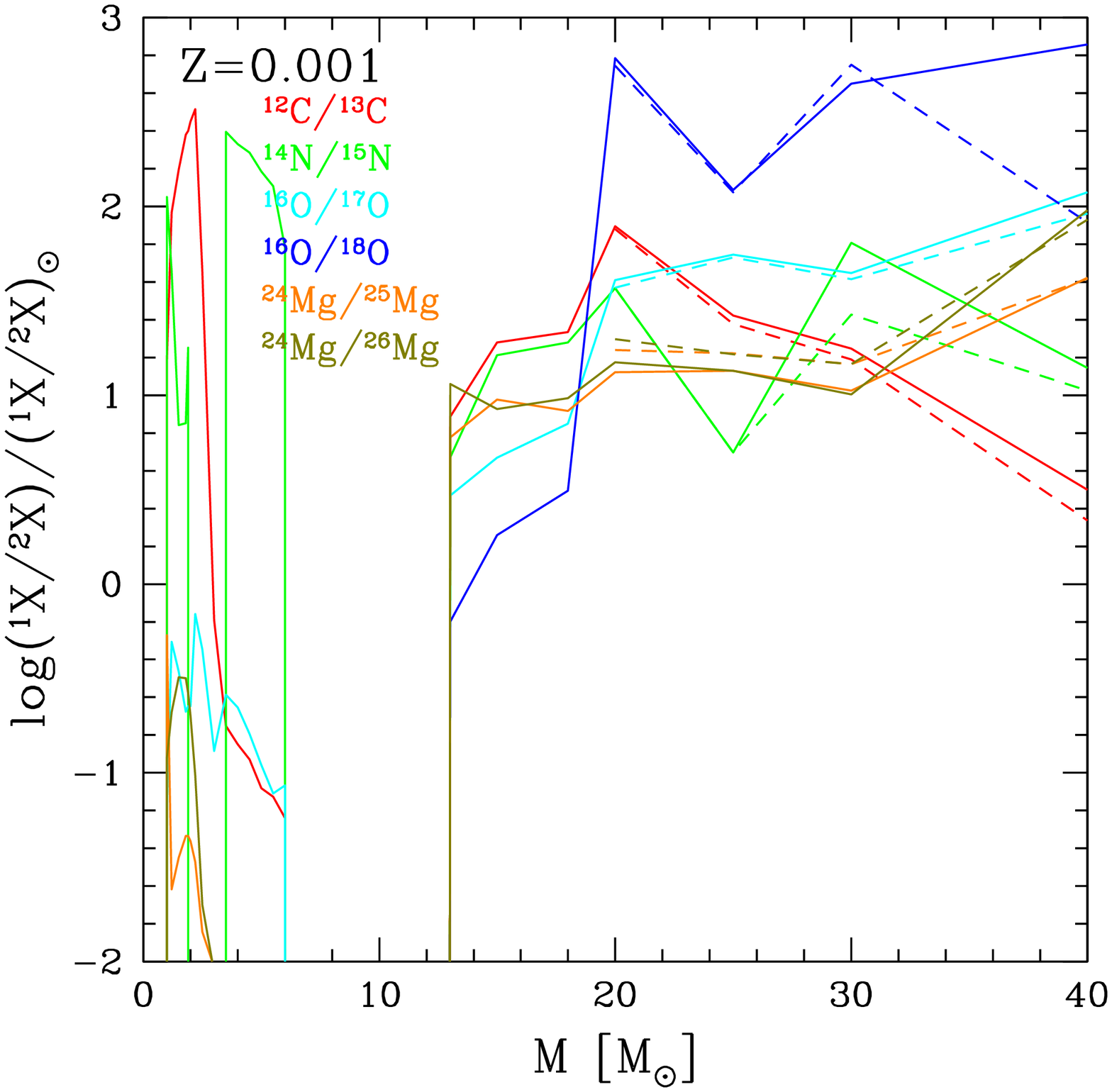}
\caption{\label{fig:isotope2}
The same as Fig.\ref{fig:isotope1} but for $Z=0.001$.
}
\end{figure}

\begin{figure}
\includegraphics[width=8.5cm]{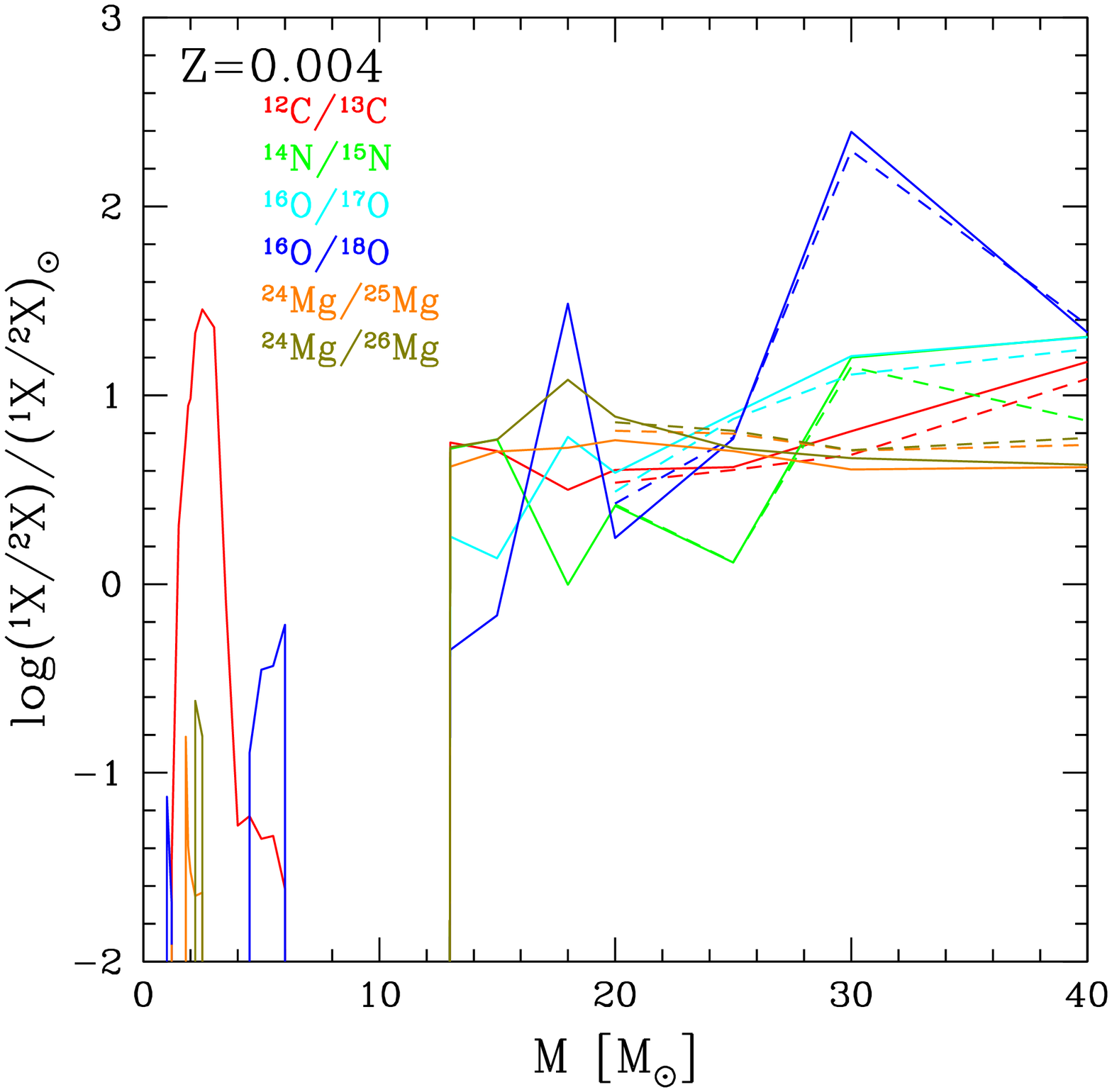}
\caption{\label{fig:isotope3}
The same as Fig.\ref{fig:isotope1} but for $Z=0.004$.
}
\end{figure}

\begin{figure}
\includegraphics[width=8.5cm]{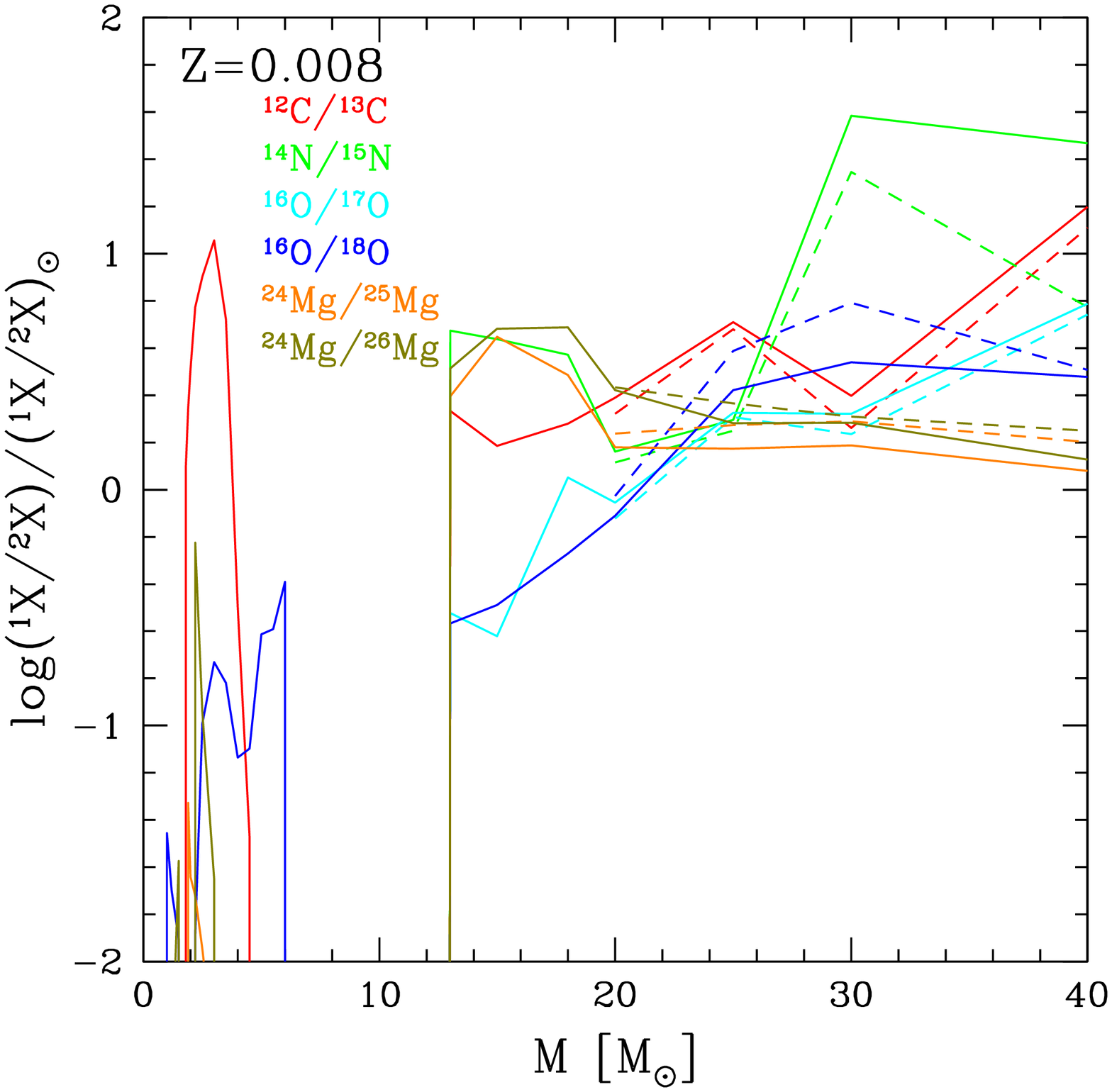}
\caption{\label{fig:isotope4}
The same as Fig.\ref{fig:isotope1} but for $Z=0.008$.
}
\end{figure}

\begin{figure}
\includegraphics[width=8.5cm]{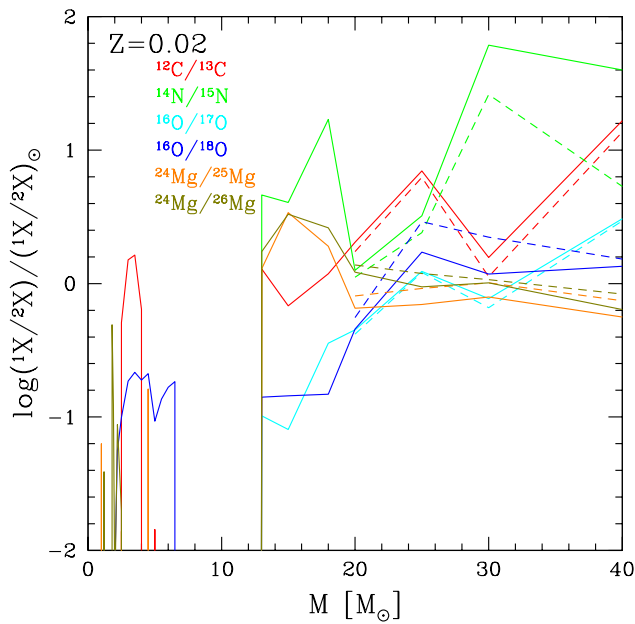}
\caption{\label{fig:isotope5}
The same as Fig.\ref{fig:isotope1} but for $Z=0.02$.
}
\end{figure}

Figures \ref{fig:yield1}-\ref{fig:yield5} and Figures \ref{fig:isotope1}-\ref{fig:isotope5} show 
the yields and the isotope ratios as a function of the mass and metallicity, normalized 
to the solar abundance \citep[hereafter AG89]{and89}.
The yields are also normalized to the mass of the ejecta $M_{\rm ejecta}$, which are much smaller 
for AGB stars than for supernovae. However, the AGB yields of C, N, and F are comparable to those of supernovae.

CNO cycling at $T\gtsim 2 \times 10^7$ K results in the production of $^{13}$C, $^{14}$N, and $^{17}$O.
Core and shell He-burning results in the synthesis of $^{12}$C, $^{16}$O, and $^{19}$F at $T\gtsim$ $1.5 \times 10^8$ K.
The heavy isotope of oxygen, $^{18}$O, is destroyed by proton captures in stellar
interiors by CNO cycling, and produced and then destroyed again by $\alpha$-captures via 
$^{14}{\rm N}(\alpha,\gamma)^{18}{\rm F}(\beta^+)^{18}{\rm O}$ followed by 
$^{18}{\rm O}(\alpha,\gamma)^{22}{\rm Ne}$ \citep[e.g., ][]{arnett1996}. 
Further secondary He-burning reactions can synthesize $^{25,26}$Mg via 
$^{22}{\rm Ne}(\alpha,n)^{25}{\rm Mg}$ and $^{22}{\rm Ne}(\alpha,\gamma)^{26}{\rm Mg}$, 
which follow after $^{18}{\rm O}(\alpha,\gamma)^{22}{\rm Ne}$.
The majority of $^{18}$O and elements heavier than fluorine up to the Fe peak are produced by hydrostatic burning 
($T\gtsim 7 \times 10^8$ K) and explosive nucleosynthesis 
($T\gtsim 2 \times 10^9$ K) in massive stars ($M \gtsim 8M_\odot$).

The AGB yields show that $^{12}$C and $^{19}$F are significantly produced in low-mass
stars ($1-4 M_\odot$) whereas  $^{13}$C, $^{14}$N, $^{25}$Mg, and
$^{26}$Mg are produced in intermediate-mass ($4-7 M_\odot$) stars.
The yields of $^{19}$F in \citet{kar10} have increased compared to \citet{kar07}, 
owing to a slower $^{19}{\rm F}(\alpha,p)^{22}{\rm Ne}$ reaction rate.
The production of F and the neutron-rich Mg isotopes in AGB stars are 
highly mass dependent. F production peaks at $\sim 3 M_\odot$ at
solar metallicity \citep{lugaro04}; in higher mass models F is
destroyed by $\alpha$-captures by the higher temperatures reached
during He-burning. Likewise, $^{25}$Mg and $^{26}$Mg are only produced
by He-burning when the temperature exceeds about $\approx 300 \times 10^{6}\,$K
and these conditions are only reached by intermediate-mass AGB stars
\citep{karakas06}. $^{17}$O is produced by the entire AGB mass range,
with relatively more from low-mass AGB stars. The yields of $^{15}$N,
$^{16}$O, and $^{18}$O are negative in most cases indicating that
these isotopes are destroyed by AGB nucleosynthesis with the largest 
destruction taking place by HBB in intermediate-mass AGB stars.

Core-collapse supernovae are the main producers of the major isotopes with more 
minor isotopes produced at higher metallicity. This is because minor isotopes are 
synthesized as secondary elements from the seed of the major isotopes.
As a result, the ratios between the major and minor isotopes are larger for supernovae 
than for AGB stars, except for $^{12}$C/$^{13}$C at $1-4 M_\odot$ (and $^{14}$N/$^{15}$N).
The ratios are, in general, larger for low-metallicity supernovae, and approach the 
solar ratios ([$^1$X/$^2$X]=0) with an increase in the metallicity.

\subsection{Initial Mass Function}

\begin{figure*}
\includegraphics[width=17.5cm]{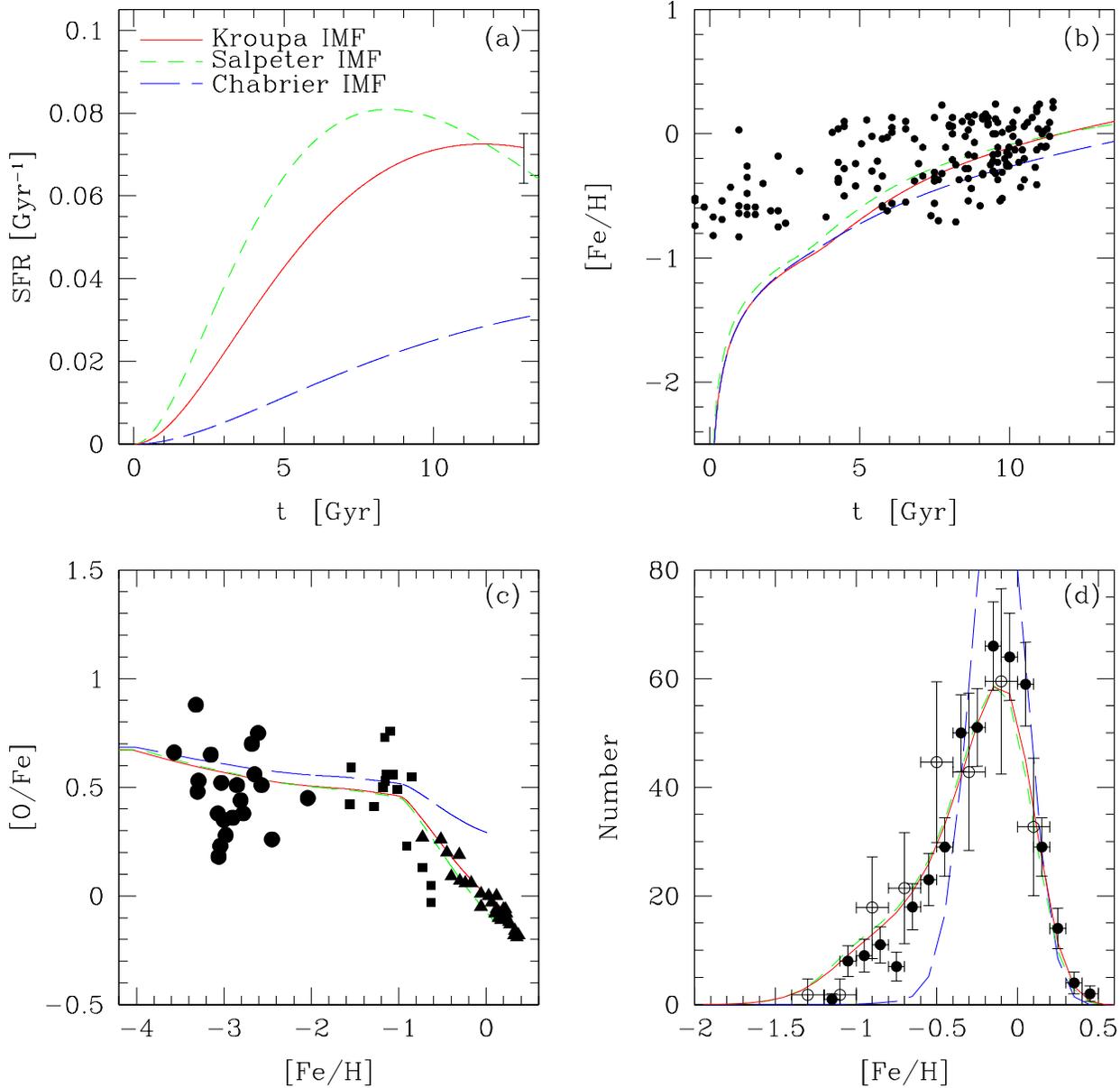}
\caption{\label{fig:imf}
Star formation histories (panel a), 
age-metallicity relations (panel b), [O/Fe]-[Fe/H] relations (panel c), and metallicity 
distribution functions (panel d) 
for the solar neighbourhood,
with the Kroupa (2008) IMF (solid lines),
the Salpeter (1955) IMF (short-dashed lines),
and the Chabrier (2003) IMF (long-dashed lines).
The observational data sources are:
an error estimate, \citet{mat97} in panel (a);
filled circles, \citet{edv93} in panels (b) and (d);
open circles, \citet{wys95} in panel (d).
In the panel (c), large filled circles, \citet{cay04}; 
filled squares, \citet{gra03};
filled triangles, \citet{ben06}.
}
\end{figure*}

We adopt the recent observational estimate of the initial mass
function (IMF) from \citet{kro07}, which is a power-law mass spectrum
$\phi(m) \propto m^{-x}$ with three slopes at different mass ranges:
$x= 1.3$ for $0.5M_\odot \le m \le 50M_\odot$,
$x= 0.3$ for $0.08M_\odot \le m \le 0.05M_\odot$, and
$x=-0.7$ for $0.01M_\odot \le m \le 0.08M_\odot$.
At the high-mass end, the slope of the Kroupa IMF is almost the same
as the \citet{sal55} IMF ($x=1.35$), and is flatter than the \citet{mil79} IMF.
Metal enrichment is only obtained from stars with $m > 0.5
M_\odot$, hence chemical evolution predictions obtained with the
Kroupa IMF are not significantly different from K06's results with 
the Salpeter IMF at $0.07M_\odot \le m \le 50M_\odot$.
In Figure \ref{fig:imf} we show that the Kroupa (short-dashed lines) 
and Salpeter (solid lines) IMFs can provide almost the same 
age-metallicity relation, [O/Fe]-[Fe/H] relation, and 
metallicity distribution function (MDF).

The mass distribution of stars from the \citet{chab03} IMF is 
peaked at $\sim 3 M_\odot$, leading to fewer stars with $m < 0.5 M_\odot$,
and a steeper slope at the high-mass end than the Kroupa and Salpeter IMFs.
Although the Chabrier IMF is adopted for many cosmological
simulations, the metal production that results is too
large to meet the observational constraints of the solar
neighbourhood. In particular the present SN Ia rate \citep{man05} 
is too high and the [$\alpha$/Fe] is too high as shown in 
Figure \ref{fig:imf} (long-dashed lines). These issues cannot 
be solved by changing SN Ia parameters.

\subsection{Star Formation Histories}

We use similar models as K06 for the star formation histories of the solar neighbourhood, bulge, halo, and thick disk, but with the Kroupa IMF.
The galactic chemical evolution is calculated with the basic equations described in K00 and K06.
In one-zone chemical evolution models, the gas fraction and the metallicity of the system evolve as a function 
of time by star formation, as well as infall and outflow of matter.
The star formation rate (SFR) is assumed to be proportional to the gas fraction ($\frac{1}{\tau_{\rm s}} f_{\rm g}$).
The infall of primordial gas from the outside of the component is given by the rate 
$\propto t\exp[-\frac{t}{\tau_{\rm i}}]$ for the solar neighbourhood, and 
$\frac{1}{\tau_{\rm i}}\exp(-\frac{t}{\tau_{\rm i}})$ for the other components.
For the halo, outflow is included such that the rate is proportional to the star formation rate 
($\frac{1}{\tau_{\rm o}}f_{\rm g}$).
For the bulge and thick disk, star formation is assumed to be truncated by galactic winds at a given epoch ($t=t_{\rm w}$).

The MDF is one of the most stringent constraints on the star formation history.
The parameters that determine the SFRs are chosen to meet the observed MDF of each component as summarized 
in Table \ref{tab:param}.
The MDFs, the resultant SFR histories, and the age-metallicity relations are shown in Figure \ref{fig:sfr}.
The uncertainties of the observations were discussed in K06 and will not be repeated in this paper.
The observed MDF in the Galactic bulge was updated by \citet{zoc08}, where the peak metallicity 
is significantly higher than the previous MDF by \citet{zoc03}.
Except for the bulge model, compared with the K06 results, the
small differences originate from the choice of the IMF and not from
the updated yields presented in this paper.

\begin{table}
\center
\caption{\label{tab:param}
Parameters of chemical evolution models: Infall, star formation, and outflow timescales, 
and the galactic wind epoch in Gyr.}
\begin{tabular}{l|cccc}
\hline
 & $\tau_{\rm i}$ & $\tau_{\rm s}$ & $\tau_{\rm o}$ & $\tau_{\rm w}$ \\
\hline
solar neighbourhood  & 5 & 4.7 & - & - \\
halo                & - & 15   & 1 & - \\
bulge               & 5 & 0.2 & - & 3 \\
thick disk          & 5 & 2.2   & - & 3 \\
\hline
\end{tabular}
\end{table}

In the solar neighbourhood model (solid lines), star formation takes place over 13 Gyr.
The MDF shows a narrow distribution peaked around [Fe/H] $\sim -0.2$, which is consistent with the observations.
Note that introducing infall significantly reduces the number of metal-poor G-dwarf stars \citep{tin80}, 
and thus there is no G-dwarf problem in the predicted MDF according to our models.

For the bulge (long-dashed lines), we use the infall$+$wind model
(model B in K06) with a short star-formation timescale.
The infall is required to explain the lack of metal-poor stars and
the wind is adopted to reproduce the sharp-cut of the observed MDF 
at the metal-rich end. It is possible that star formation continues 
at present time, forming super metal-rich stars in the Galactic bulge.
In this case the SFR should be low to meet the MDF. 
In our bulge model the duration of star formation is set to be 3 Gyr, 
which results in a peak metallicity of [Fe/H] $\sim +0.3$. 
A much higher efficiency of chemical enrichment, e.g., a flatter IMF 
is not required, unless the duration is much shorter than 3 Gyr.
Note that the 3 Gyr duration is consistent with chemodynamical simulations of Milky Way-type galaxies from CDM initial conditions (KN11).

For the thick disk (dot-dashed lines), we use the infall$+$wind model (model C in K06), which also gives 
a good agreement with the observed age-metallicity relation \citep{ben04b}.
In the thick disk model, the formation timescale is as short as $\sim 3$ Gyr, and the star formation 
efficiency is larger than that for the solar neighbourhood but smaller than for the bulge.
The assumption that the timescale of star formation is shorter in the 
bulge and thick disk was suggested by \citet{mat90}, but by itself does not completely 
explain the observed abundance patterns.  A more intense star formation is also required as well 
as a shorter star formation timescale.

For the halo (short-dashed lines), we use an outflow model without infall, which results in a low mean metallicity 
of [Fe/H] $\sim -1.6$ \citep[e.g.,][]{chi98}.
In the halo model, the star formation efficiency is much lower than for the other components, 
and the outflow causes an effective metal loss.

The metallicity dependent main-sequence lifetimes are taken from \citet{kod97} for $0.6-80M_\odot$, 
which are calculated with the stellar evolution code described in \citet{iwan99}.
These are in excellent agreement with the lifetimes in \citet{kar10} for low- and intermediate-mass stars.

\section{Results}

\begin{figure*}
\includegraphics[width=17.5cm]{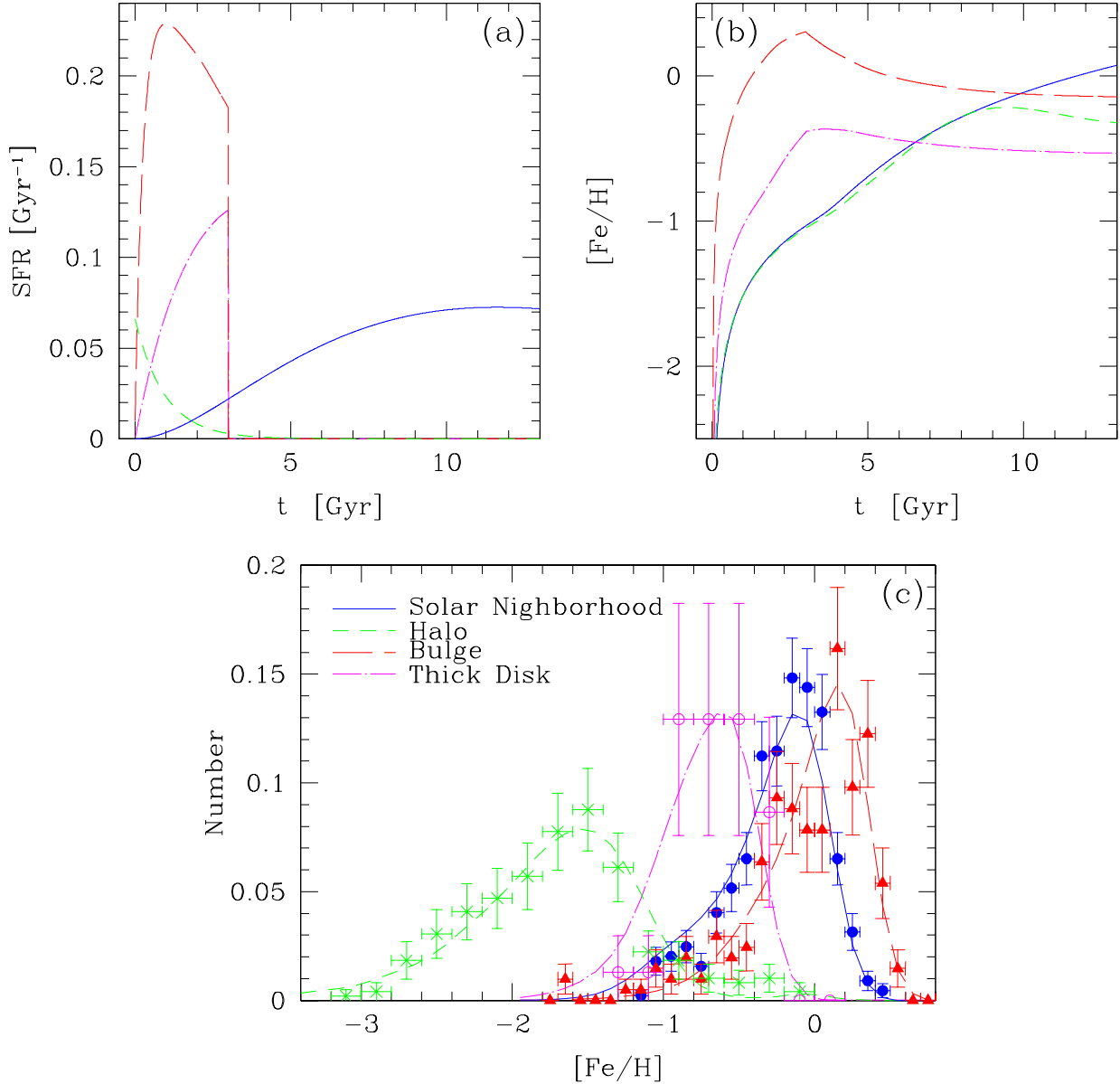}
\caption{\label{fig:sfr}
Star formation histories (panel a), age-metallicity relations (panel b), and metallicity 
distribution functions (panel c) for the 
solar neighbourhood (solid lines),
halo (short-dashed lines),
bulge (long-dashed lines),
and thick disk (dot-dashed lines).
The observational data sources are:
filled circles, \citet{edv93};
crosses, \citet{chi98};
filled triangles, \citet{zoc08};
open circles, \citet{wys95}.
}
\end{figure*}

\begin{figure*}
\includegraphics[width=22.7cm]{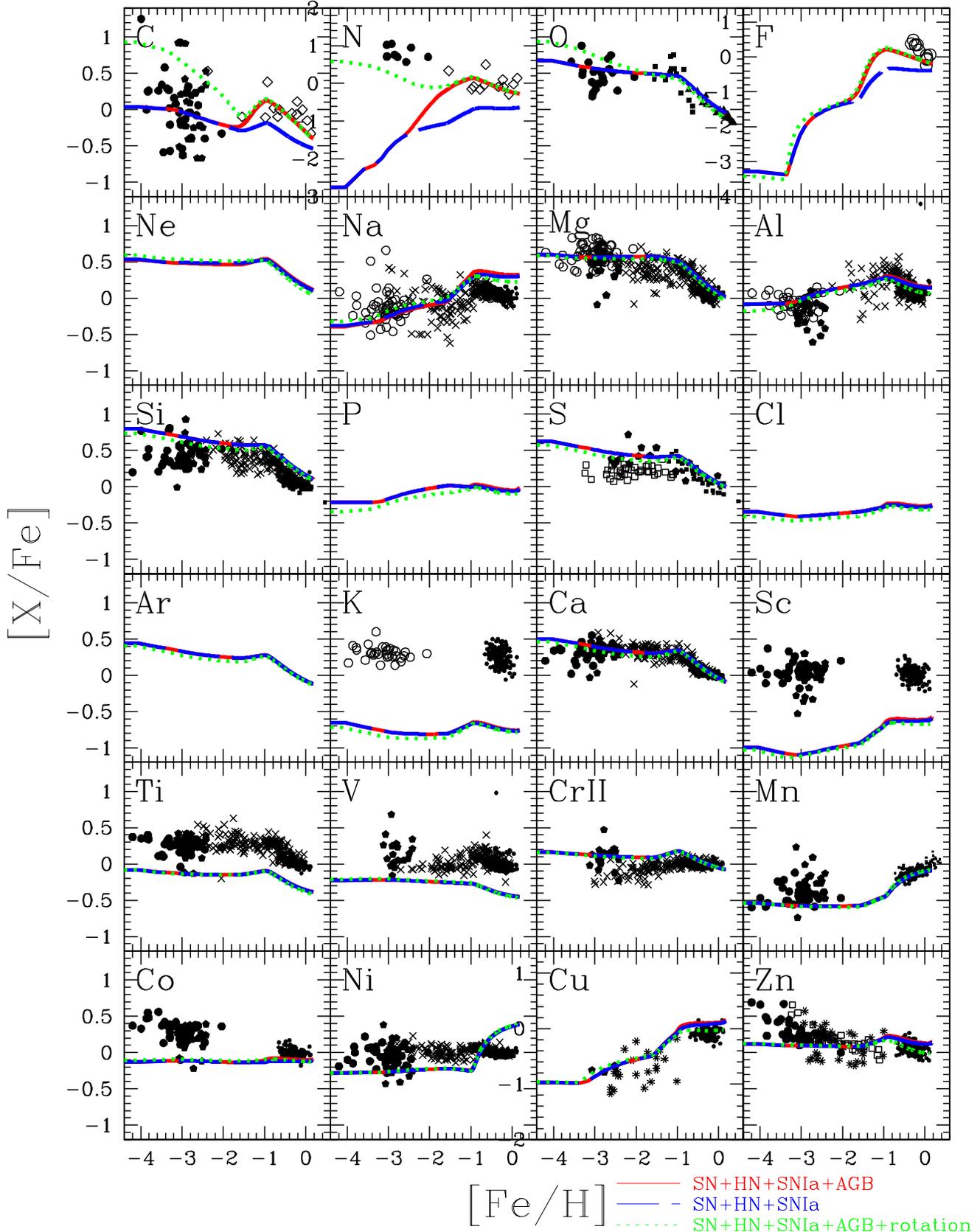}
\caption{\label{fig:xfe}
Evolution of elemental abundance ratios [X/Fe] against [Fe/H]
for the solar neighbourhood
with SNe II, HNe, and SNe Ia only (dashed lines), 
with AGB stars (solid lines),
and with rotating massive stars at $Z=0$ (dotted lines).
The dots are observational data (see KN11 for the references).
For C and N, only unevolved stars are plotted.
}
\end{figure*}

\begin{figure*}
\includegraphics[width=22.7cm]{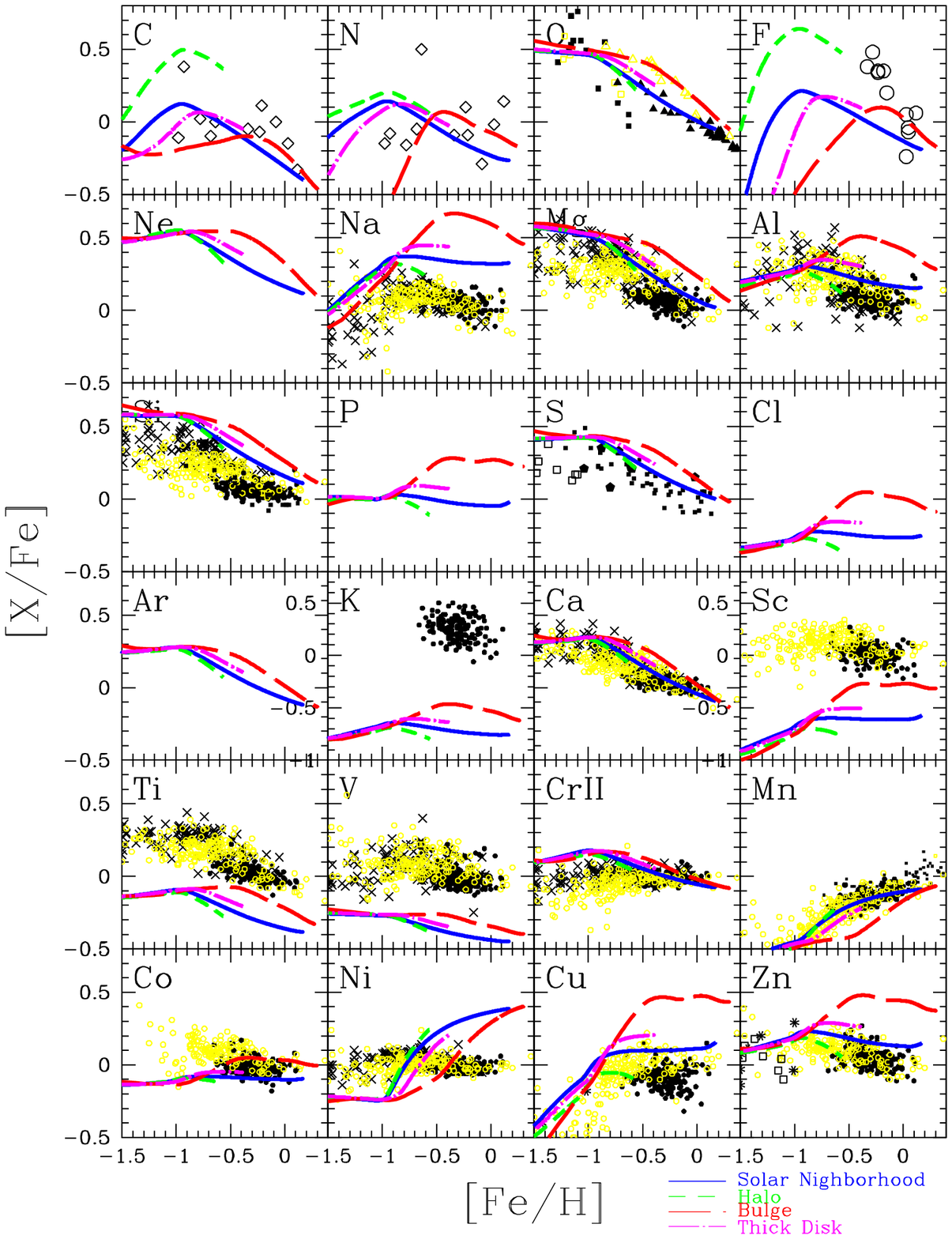}
\caption{\label{fig:xfe2}
Evolution of elemental abundance ratios [X/Fe] against [Fe/H]
for the solar neighbourhood (solid lines),
halo (short-dashed lines),
bulge (long-dashed lines),
and thick disk (dot-dashed lines)
with AGB yields.
The black and yellow dots are observational data for thin and thick disk stars (see KN11 for the references).
}
\end{figure*}

Figure \ref{fig:xfe} shows the 
evolution of element abundance ratios [X/Fe] against [Fe/H]
for the solar neighbourhood, using the models
with SNe II, HNe, and SNe Ia only (dashed lines), 
with AGB stars (solid lines),
and with rotating massive stars at $Z=0$ (dotted lines).
The difference among models is seen only for C, N, and F.
For the elements heavier than Na,
the small difference among the models with and without rotating 
massive stars is caused by the difference in the IMF ($M_{\rm u,2}=120M_\odot$ is adopted in 
the case with rotating massive stars instead of $M_{\rm u,2}=50M_\odot$).
Observational data are taken from several sources, which were selected to minimize systematic errors
as discussed in KN11. In that paper, we also discussed the different results for the lines used in the 
observational data analysis, which will not be repeated in this paper. For C and N, only data from 
unevolved stars are plotted.
Both in the models and the observational data, the solar abundances from AG89 are adopted.
The results are summarized as follows.

\begin{itemize}
\item {$\alpha$ elements} ---
In the early stages of galaxy formation only SNe II/HNe contribute and the [$\alpha$/Fe] ratio quickly reaches 
a plateau ([$\alpha$/Fe] $\sim 0.5$).
Around [Fe/H] $\sim -1$ SNe Ia start to occur, which produce more iron than $\alpha$ elements. 
This delayed enrichment of SNe Ia causes the decrease in [$\alpha$/Fe] with increasing [Fe/H].
The [Fe/H] where the [$\alpha$/Fe] starts to decrease depends on the adopted SN Ia progenitor model, 
and is determined not by the lifetime but by the metallicity dependence of SN Ia progenitors (KN09).
As a result, this trend is in excellent agreement with the observations (dots) for O, Mg, Si, S, and Ca. 
Ne and Ar also show a similar trend.
Ti is underabundant overall, but this problem can be solved with 2D nucleosynthesis calculations
\citep{mae03}, as well as for Sc and V.
AGB stars do not make any difference to these trends.
Although AGB stars produce significant amounts of the Mg
isotopes, the inclusion of these do not affect the [Mg/Fe]-[Fe/H] relation.

\item {Odd-Z elements} ---
Na, Al, and Cu show a decreasing trend toward lower metallicity, which is well reproduced 
by the strong dependence of these elements on the metallicity of the progenitor stars (see Fig.5 in K06).
In contrast, Na and Al show a decreasing trend toward higher metallicity owing to the contribution from SNe Ia, 
which is shallower than the trend for the $\alpha$ elements.
Such a decrease is not seen for Cu since Cu is also produced by SNe Ia.
With the updated AGB yields (\S 2.2), [Na/Fe] is consistent with the observations, and the Na overproduction 
problem by AGB stars is not seen. Note that AGB stars may produce some Cu but no yields are available for 
a large range of masses and metallicities.
K, Sc, and V are underabundant overall, a problem which has not been discussed in detail in previous studies. 
The $\nu$-process can increase the production of these elements (Izutani, Umeda, \& Yoshida 2010, 
private communication), although the yields are not yet available.
[(P, Cl)/Fe] are also negative overall in our predictions.
There is a metallicity dependence of P, Cl, K, and Sc yields at $Z>0.001$ for SNe II/HNe, which 
causes a weak decrease from [Fe/H] $\sim -1$ to $\sim -3$.
The V yields do not depend very much on metallicity.

\item {Iron-peak elements} ---
[(Cr, Mn, Co, Ni, Zn)/Fe] are consistent with the observed mean values at $-2.5 \ltsim$ [Fe/H] $\ltsim -1$.
Note that Cr II observations are plotted, because this line is not strongly affected by NLTE effect.
For Mn, the NLTE effect should not be so large as indicated by Mn II observations \citep{joh02,mas10}, 
although a strong NLTE effect is reported by \citet{ber08}.
At [Fe/H] $\ltsim -2.5$, observational data show an increasing trend of [(Co, Zn)/Fe] toward lower metallicity, 
which will not discussed here since inhomogeneous chemical enrichment is becoming increasingly important.
The [(Co, Zn)/Fe] trend can be explained by HNe under the assumption that the observed stars were enriched by only a single supernova \citep{tom07}.

\item {Manganese} ---
Mn is a characteristic element of SN Ia enrichment and is produced more by SNe Ia than SNe II/HNe relative to iron.
From [Fe/H] $\sim -1$, [Mn/Fe] shows an increasing trend toward higher metallicities, which is caused by the 
delayed enrichment of SNe Ia.
\citet{fel07} showed a steep slope at [Fe/H] $>0$,
which could be generated by the metallicity dependence of SN Ia yields.
\citet{ces08} demonstrated this by applying an artificial metallicity dependence to the theoretical calculations.
In principle, Mn is an odd-Z element and the Mn yields depend on the metallicity both of SNe II and SNe Ia.
The metallicity dependence for SNe II are included in our yields. For SNe Ia, a strong metallicity dependence 
is not expected (H. Umeda, K. Nomoto, et al., private communication).

\item {Zinc} ---
Zn is one of the most important elements for supernova physics.
[Zn/Fe] is about $\sim 0$ for a wide range of metallicities, which can only be generated by a large 
fraction of HNe (50\% of $M\ge20M_\odot$).
In detail, there is a small oscillating trend; [Zn/Fe] is $0$ at [Fe/H] $\sim 0$, this increases to $0.2$ 
at [Fe/H] $\sim -0.5$, decreases to be $0$ again at [Fe/H] $\sim -2$, then increases toward lower metallicity.
This is characteristic of our SN Ia model (KN09) and is consistent with the observations \citep{sai09}.
Theoretically, Zn production depends on many parameters;
$^{64}$Zn is synthesized in the deepest region of HNe, while neutron-rich isotopes of zinc $^{66-70}$Zn 
are produced by neutron-capture processes, which are larger for higher metallicity massive SNe II.
Since the observed [Zn/Fe] ratios show an increasing trend toward lower metallicity \citep{pri00,nis07,sai09}, 
the HN fraction may have been larger in the earliest stages of galaxy formation.
At higher metallicities, the HN fraction may be as small as $1\%$ (KN11).

\item {Carbon} ---
Although the ejected mass of C is similar for low-mass AGB ($1-4 M_\odot$) and massive ($>10 M_\odot$) 
stars, the [C/Fe] ratio is enhanced efficiently by low-mass stars because these stars produce no Fe
(Figs. \ref{fig:yield1}-\ref{fig:yield5}).
When we include AGB yields (solid lines), [C/Fe] increases from [Fe/H] $\sim -1.5$, which corresponds to 
the lifetime of $\sim 4 M_\odot$ stars ($\sim 0.1$ Gyr).
At [Fe/H] $\sim -1$, [C/Fe] reaches $0.13$, which is $0.32$ dex larger than the case without AGB yields (dashed lines).
This is roughly consistent with previous models such as \citet{pra94}.
At [Fe/H] $\gtsim -1$, [C/Fe] shows a decrease due to SNe Ia.
Because of the long lifetimes of AGB stars, no difference is seen at [Fe/H] $\ltsim -1.5$, which is consistent 
with the observed behavior of s-process elements \citep{travaglio04}.
If we include the yields of rotating massive stars (dotted lines), the [C/Fe] ratio becomes as large as 
$\sim 0.5$ at [Fe/H] $\sim -2.5$.
A significant fraction of metal-poor stars show carbon enrichment (CEMP stars), with several scenarios
proposed to explain the observed abundances including a single supernova \citep[e.g.,][]{ume02} and AGB stars in 
binary systems \citep{sud04,lugaro08,izz09}. Such local peculiar effects are not included in our models.
We should also note that AGB stars can contribute at metallicities below [Fe/H] $\ltsim -1.5$ when
an inhomogeneous chemical enrichment is taken into account (KN11).

\item {Nitrogen} ---
Different from C, N is produced mainly by intermediate-mass AGB stars ($4-7 M_\odot$,
independent of the integration on the IMF).
Therefore, the contribution from AGB stars (solid lines) is seen at [Fe/H] $\sim -2.5$.
At [Fe/H] $\sim -1$, [N/Fe] reaches $0.15$, which is $0.8$ dex larger than the case without AGB yields (dashed lines).
At [Fe/H] $\gtsim -1$, [N/Fe] shows a shallow decrease due to SNe Ia.
No difference is seen at [Fe/H] $\ltsim -2.5$ with and without the AGB yields, while [N/Fe] can be as large as 
$\sim 0.5$ with rotating massive stars (dotted lines).
\citet{chi06} showed that the contribution from rotating massive stars is required to solve the primary N problem.
As noted above, however, AGB stars can also contribute to N production even at [Fe/H] $\ltsim -2.5$ 
when taking inhomogeneous chemical enrichment into account (KN11).
From the difference between C and N, it is possible to distinguish the contribution from low- and 
intermediate-mass AGB stars as the enrichment source of the observed metal-poor stars.
\citet{pol09} showed that the IMF with the Gaussian distribution peaked at $\sim 10 M_\odot$ \citep{kom07} 
is rejected when trying to match the fraction between C-rich and N-rich stars, based on binary population 
synthesis models (see also \citealt{izz09}).

\item {Fluorine} ---
F is one of the most interesting elements, although F abundances are estimated from only one infrared line 
from stellar spectra. AGB stars and massive stars have both been suggested to produce F but production has
only been confirmed for AGB stars \citep{jor92,abia10}. The AGB mass range that produces F is similar for C and 
is $2-4 M_\odot$.  Thus the difference is seen only at [Fe/H] $\gtsim -1.5$.
At [Fe/H] $\sim -1$, [F/Fe] reaches $0.22$ in the model with the AGB yields (solid lines). 
This is $0.56$ dex larger than the case without the AGB yields (dashed lines) and much closer to the 
observational data \citep{cun03}.
Note that the F yields from AGB stars were increased with the new reaction rates (\S 2).
Different from C, F is not significantly produced by SNe II/HNe according to our yields, and 
thus [F/Fe] rapidly decreases from [Fe/H] $\sim -1$ to $\sim -3$.
Therefore, the F abundance is a good clock to distinguish the contribution from low-mass AGB 
stars and supernovae. 
We should note, however, that the F yields from supernovae may be increased by a factor of $\sim 1000$ by the 
$\nu$-process (Izutani, Umeda, \& Yoshida 2010, private communication; see also \citealt{woo95}).
The effect of rotating massive stars is uncertain since F yields are not available in the literature.

\end{itemize}

\begin{figure}
\includegraphics[width=8.5cm]{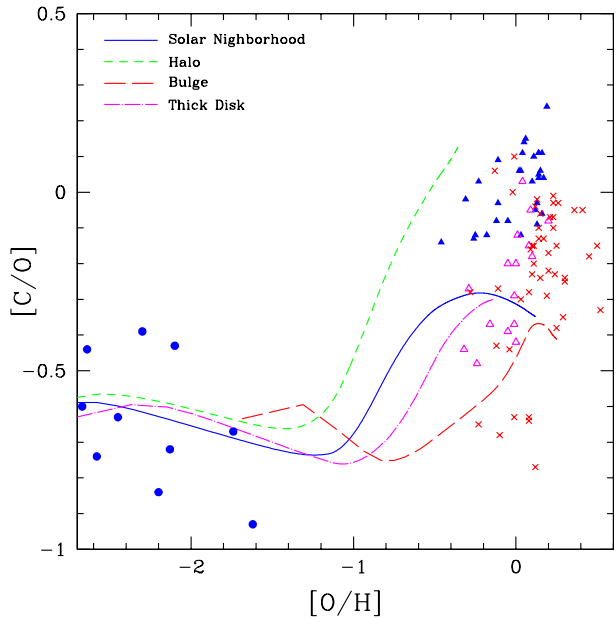}
\caption{\label{fig:co}
Evolution of the [C/O] ratio against [O/H]
for the solar neighbourhood (solid lines),
halo (short-dashed lines),
bulge (long-dashed lines),
and thick disk (dot-dashed lines)
with AGB yields.
The observational data sources are:
filled circles, unmixed stars in \citet{spi06};
filled triangles and open triangles, \citet{ben06} for thin and thick disk stars, respectively;
crosses, \citet{lec07} for bulge stars.
}
\end{figure}

\begin{figure}
\includegraphics[width=8.5cm]{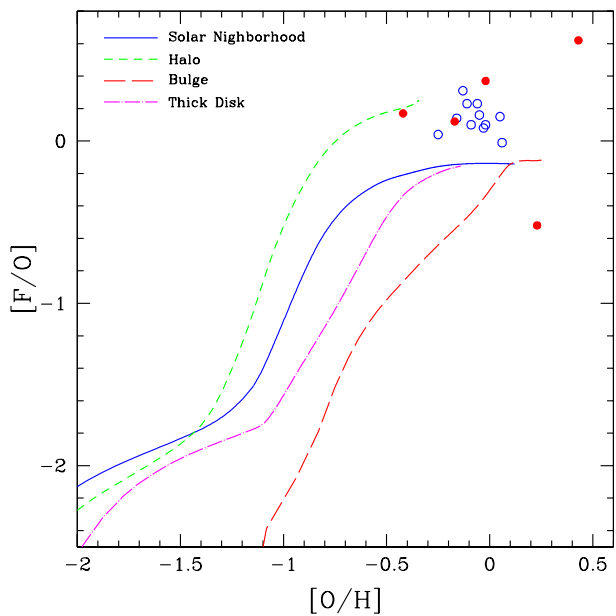}
\caption{\label{fig:fo}
The same as Fig. \ref{fig:co} but for the [F/O] ratio.
The observational data sources are:
open circles, \citet{cun03} for the solar neighbourhood stars;
filled circles, \citet{cun08} for bulge stars.
}
\end{figure}

These results are consistent with the models that adopt the same 
supernova and AGB yields in \citet[hereafter R10]{rom10}.
There are some differences, for example, for 
elements heavier than N, our results fall between Models 4 and 5 in 
R10 because we adopt an hypernova efficient of $\epsilon_{\rm HN}=0.5$.
The C and N abundances are predicted to be higher in Models 4 and 5 than our results because the
low and intermediate-mass star yields of \citet{van97} were used. 
In Model 15 of R10, the same AGB yields are adopted as in our models, but the C and N abundances are 
still higher. This may be due to the addition of the Geneva pre-supernova yields along with the supernova yields
from K06. We note that rotation changes the pre-supernova structure and thus should also change the
nucleosynthesis during the explosion. Hence to be fully self consistent, supernovae yields computed from 
a rotating pre-supernova structure should be included but these are not yet available for a wide range of
masses and metallicities.
Other model differences include the adopted IMF, star formation rates, and the SN Ia model, but these 
do not significantly affect the average evolution of elemental abundance ratios.
However, we do observe small differences in e.g., the [$\alpha$/Fe]-[Fe/H] relations as a result of the
the different adopted SN Ia model. In R10's models, the evolutionary change in [O/Fe] around [Fe/H] 
$\sim -1$ is not as sharp as in our models. In addition, the evolutionary track around [Fe/H] $\sim -0.8$ 
in the R10 models is not smooth, which may result in a large number of stars at this point in 
the [O/Fe]-[Fe/H] diagram.
Large and homogeneous observational data sets may help to put constraints on the modelling of SNe Ia.

Enrichment sources produce different elements on different timescales, and thus the time evolution of the elements varies as a function of location in a galaxy, depending on the star formation history.
In Figure \ref{fig:xfe2} we show the evolution of elemental abundance ratios [X/Fe] against [Fe/H]
for the solar neighbourhood (solid lines),
halo (short-dashed lines),
bulge (long-dashed lines),
and thick disk (dot-dashed lines),
where the contributions from SNe II, HNe, SNe Ia, and AGB stars are included.

{\bf Bulge and thick-disk ---}
If the star formation timescale is shorter than in the solar neighbourhood (solid lines) as in our bulge 
(long-dashed lines) and thick-disk (dot-dashed lines) models, the contribution from stars of a given 
lifetime appear at a higher metallicity than in the solar neighbourhood.
Intermediate-mass AGB stars, low-mass AGB stars, and SNe Ia start to contribute at [Fe/H] $\sim -2.5, -1.5$, and $-1$, 
respectively in the solar neighbourhood, but at a higher [Fe/H] in the bulge and thick disk models.
At [Fe/H] $\gtsim -1$, [$\alpha$/Fe] is higher and [Mn/Fe] is lower than in the solar neighbourhood 
because the SN Ia contribution is smaller in the bulge and thick disk.
Simultaneously, the [(C, N, F)/Fe] ratios peak at higher metallicities; [(C, N, F)/Fe] is lower 
at [Fe/H] $\ltsim -1$, and is slightly higher at [Fe/H] $\gtsim -0.5$ than in the solar neighbourhood.
The abundance ratios of [(Na, Al, Cu, Zn)/Fe] and [(P, Cl, K, Sc)/Fe] are predicted to be higher because 
of the higher metallicity. Indeed, the metallicity reaches values high enough to produce these elements 
before the majority of SNe Ia occur.
The yellow dots are the observational data of the thick disk stars, which show higher [$\alpha$/Fe] 
and [(Al, Cu)/Fe] ratios than the thin disk stars (black dots). These data are roughly consistent with our 
model predictions.
Note that the transition metallicity where [$\alpha$/Fe] starts to decrease and the peak metallicity of 
[(C, N, F)/Fe] and [(Na, Al, Cu, Zn)/Fe] depends on the star formation and infall timescales of the system. 
For the bulge model, [O/Fe] starts to decrease at [Fe/H] $\sim -0.5$. The transition metallicity can 
be increased with a much shorter timescale as in \citet{bal07}.

The high [$\alpha$/Fe] in the bulge and thick disk can also be
reproduced by changing the IMF, namely, adopting a flatter IMF (see
Fig. 32 of K06). In this paper we do not need to change the IMF
between the solar neighbourhood and the bulge (\S 2.4). This is because the 
observational constraints of the MDF was updated by \citet{zoc08} 
and the bulge stars are more metal-rich than in the solar
neighbourhood (Fig.\ref{fig:sfr}). With a flatter IMF, the predicted 
[(Zn,Co)/Fe] ratios becomes much larger, a result that can be 
tested with future observations.

\begin{figure*}
\vspace*{-5cm}
\includegraphics[width=17.5cm]{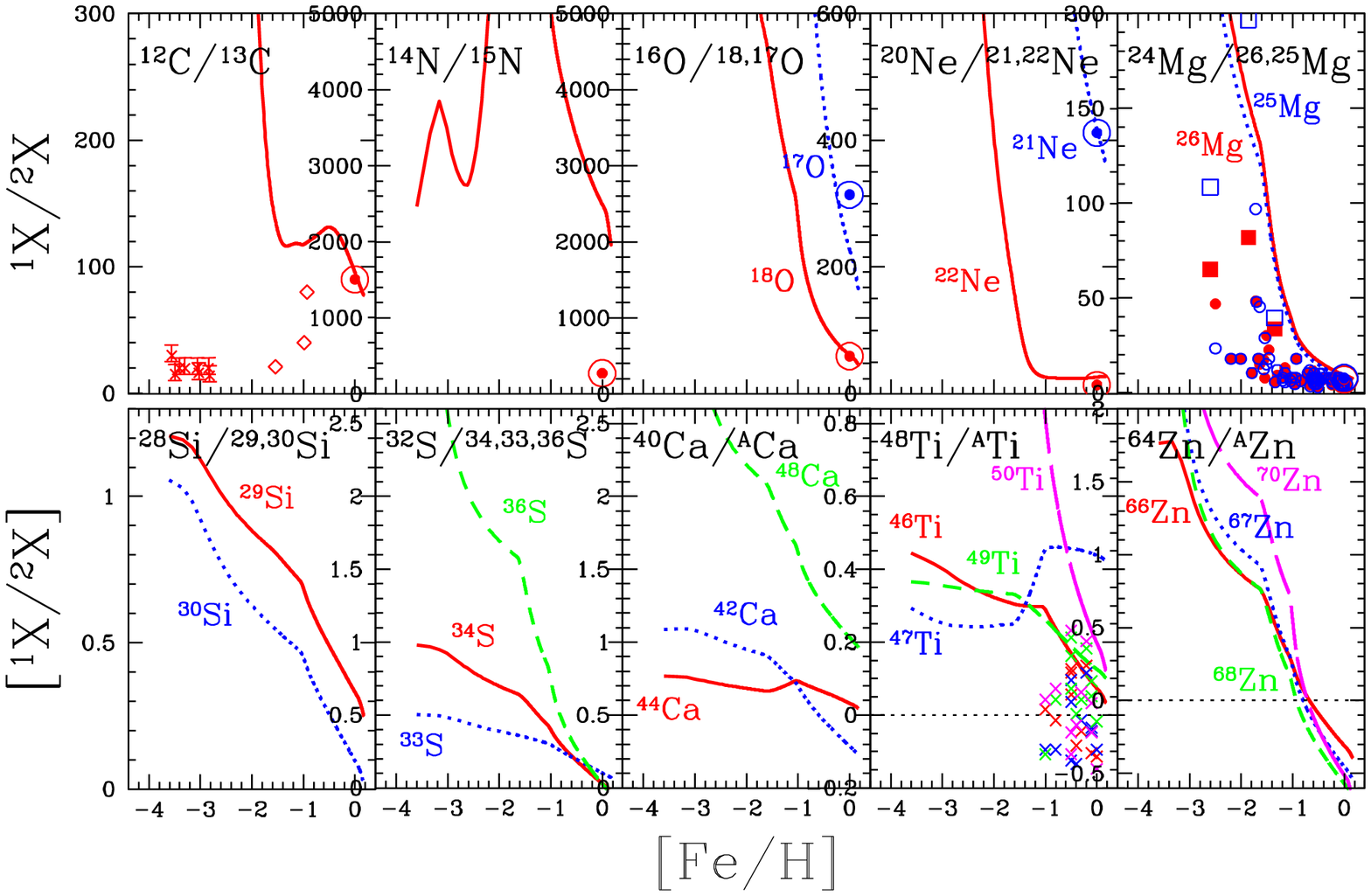}
\caption{\label{fig:iso}
Evolution of isotope ratios against [Fe/H]
for the solar neighbourhood (solid lines)
with AGB yields.
Observational data sources include:
For C,
\citet{car00}, diamonds;
\citet{spi06}, asterisks;
For Mg,
\citet{yon03}, open and filled circles for $^{25}$Mg and $^{26}$Mg;
\citet{mel07}, open and filled squares for $^{25}$Mg and $^{26}$Mg;
For Ti,
\citet{cha09}, crosses.
The solar ratios \citep{and89} are shown with the solar symbols at [Fe/H] $=0$ in the upper panels.
}
\end{figure*}

\begin{figure*}
\vspace*{-5cm}
\includegraphics[width=17.5cm]{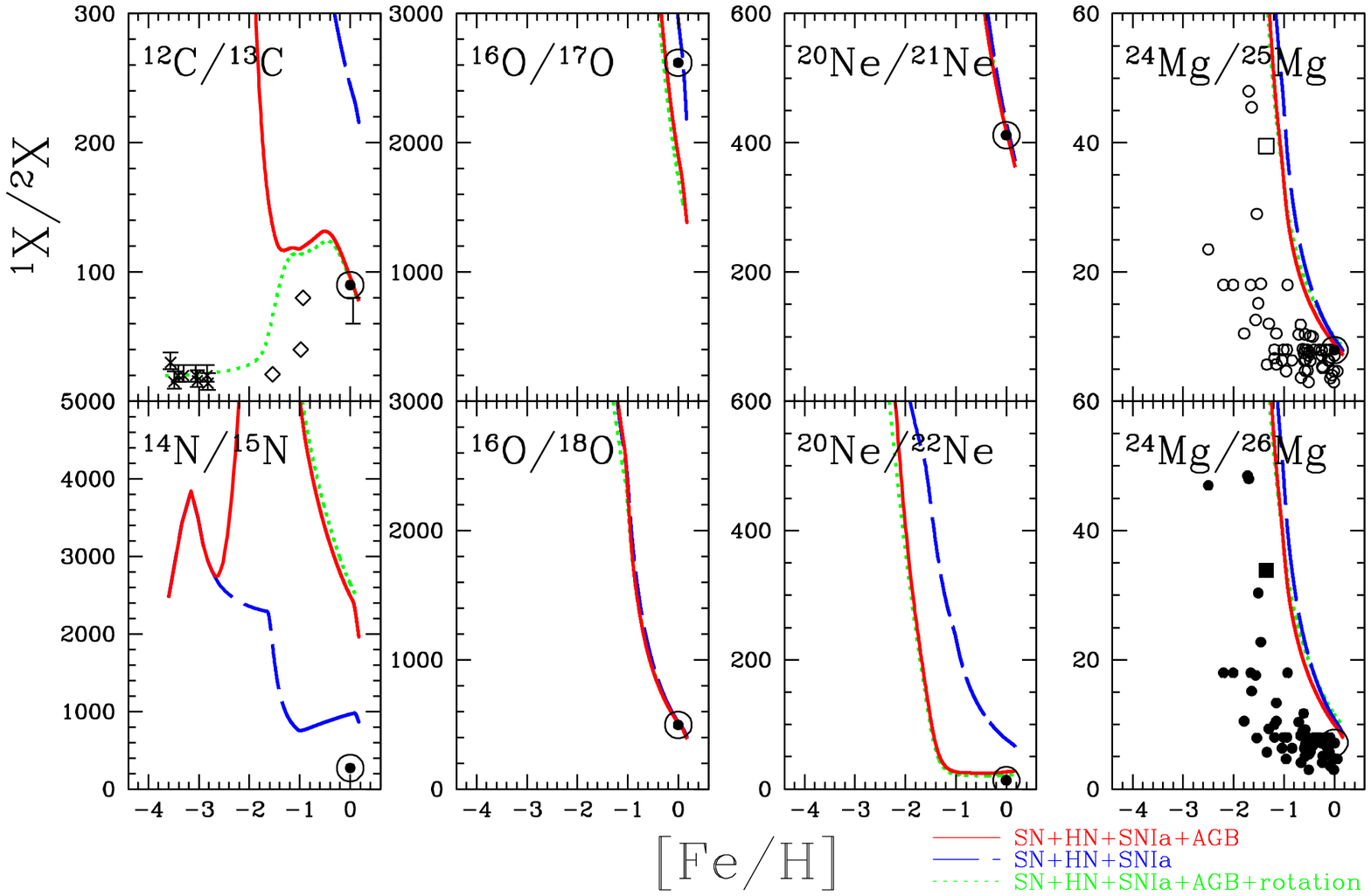}
\caption{\label{fig:iso2}
Evolution of isotope ratios against [Fe/H]
for the solar neighbourhood
with SNe II, HNe, and SNe Ia only (dashed lines), 
with AGB stars (solid lines),
and with rotating massive stars at $Z=0$ (dotted lines).
See Fig.\ref{fig:iso} for the observational data sources.
}
\end{figure*}

The difference in the AGB contribution is clearly seen in [(C, N, F)/O] ratios.
Figure \ref{fig:co} shows the evolution of [C/O] against [O/H], where [C/O] shows a rapid increase 
at [O/H] $\sim -1$ to $\sim -0.5$ toward higher metallicities.
At [O/H] $\gtsim -1$, the [C/O] ratio is highest in the halo, followed by the solar neighbourhood, 
thick disk, and bulge.
This is qualitatively consistent with the observations (dots) in the thin disk, thick disk, and bulge, but the very high [C/O] ratios that are observed in some stars at [O/H] $\sim 0$ cannot be reproduced by our models.
\citet{ces09} showed that the [C/O] ratios could be enhanced at [O/H] $\gtsim 0$ if \citet{mae92} 
or \citet{mey02} yields are included.
Similarly, Figure \ref{fig:fo} shows the evolution of [F/O] against [O/H], where [F/O] shows a 
rapid increase at [O/H] $\sim -1.5$ to $\sim -1$ toward higher metallicities.
The present [F/O] ratio is slightly lower than the observations, perhaps indicating the need for other
sources of F in the galaxy. \citet{ren04} showed that [F/O] could be enhanced at [O/H] $\gtsim -0.2$ 
if the yields of Wolf-Rayet stars are included.

\citet{mcw08} showed that the [O/Mg] ratio could be changed in their bulge models if the 
\citet{mae92}'s yields are included. This ratio cannot be changed by the addition of AGB yields.
With our yields, [O/Mg] is almost constant independent of the metallicity as shown in Fig. 9 of K06.
Our model predictions are consistent with the observations of thin disk stars, but not with those 
of thick disk and bulge stars, although the scatter and uncertainties are quite large for the bulge stars.

{\bf Halo ---}
If the chemical enrichment timescale is longer than in the solar neighbourhood as in our halo model 
(short-dashed lines), the contribution from low-mass AGB stars become significant, and thus the 
[(C, F)/Fe] ratios are higher at all metallicities than in the solar neighbourhood.
For [N/Fe], the difference is seen only at low metallicity ([Fe/H] $\ltsim -1$) since the mass range 
of N production is more massive than that of C and F.
The [$\alpha$/Fe] and [Mn/Fe] relations are the same as in the solar neighbourhood.
However, the [(Na, Al, Cu, Zn)/Fe] and [(P, Cl, K, Sc)/Fe] ratios are lower because of the 
overall lower metallicity.
Detailed elemental abundances are not available for the Galactic halo, but high [C/Fe] is seen 
for a significant fraction of stars in the Galactic outer halo in  the SEGUE\footnote{Sloan Extension for 
Galactic Understanding and Exploration} data \citep{bee10}.

\begin{figure*}
\vspace*{-5cm}
\includegraphics[width=17.5cm]{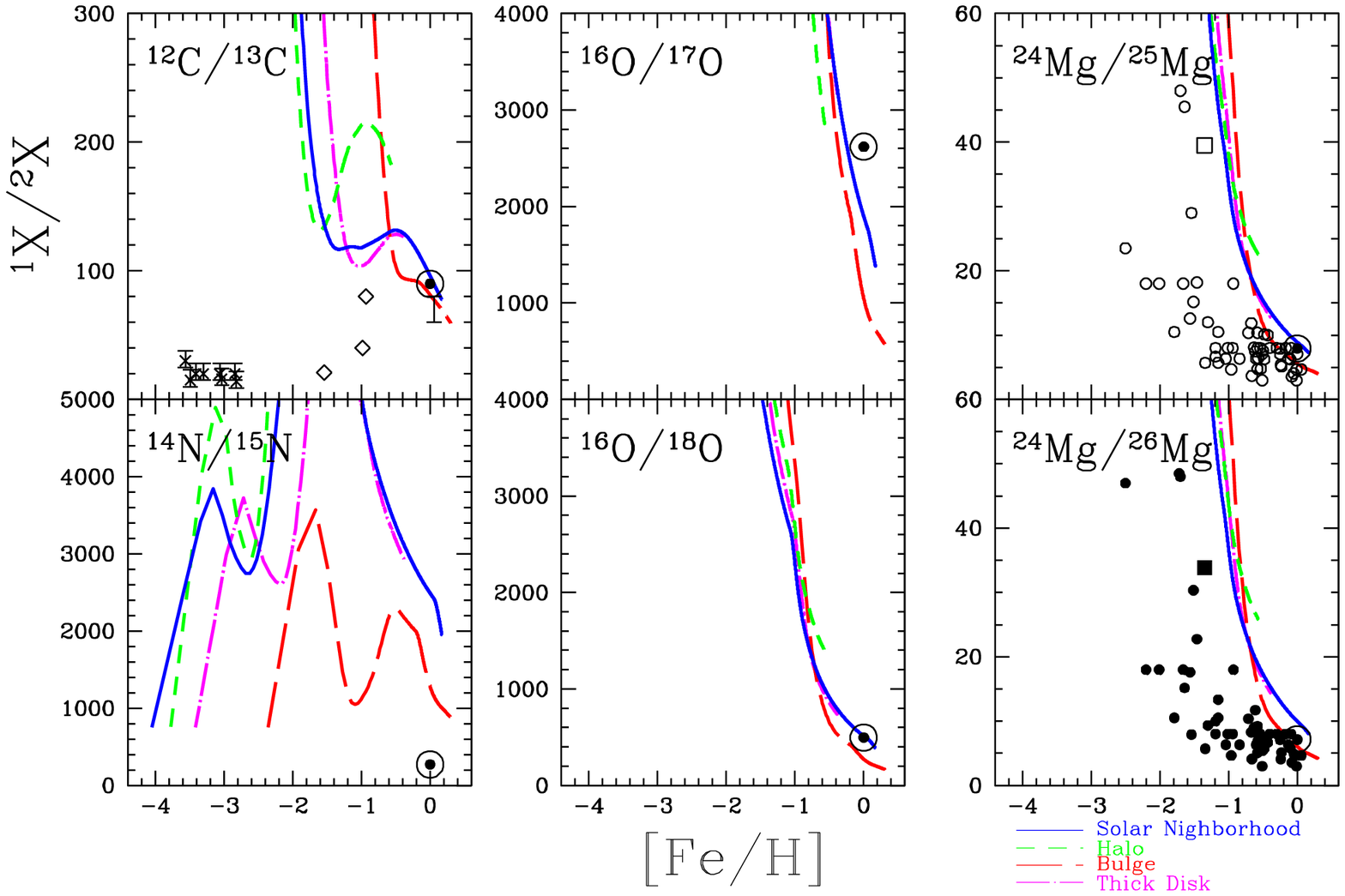}
\caption{\label{fig:iso3}
Evolution of isotope ratios against [Fe/H]
for the solar neighbourhood (solid lines),
halo (short-dashed lines),
bulge (long-dashed lines),
and thick disk (dot-dashed lines)
with AGB yields.
See Fig.\ref{fig:iso} for the observational data sources.
}
\end{figure*}

Figure \ref{fig:iso} shows the 
evolution of the isotope ratios against [Fe/H]
for the solar neighbourhood (solid lines).
In general, core-collapse supernovae are the main producers of the major isotopes with more 
minor isotopes synthesized at higher metallicity (see Figs. \ref{fig:isotope1}-\ref{fig:isotope5}).
For this reason, the evolution of $^{12}$C/$^{13}$C, $^{16}$O/$^{17,18}$O, $^{20}$Ne/$^{21,22}$Ne,  
$^{24}$Mg/$^{25,26}$Mg, $^{28}$Si/$^{29,30}$Si, $^{32}$S/$^{33,34,26}$S, $^{40}$Ca/$^{42,44,48}$Ca, 
$^{48}$Ti/$^{46,47,49,50}$Ti, and $^{64}$Zn/$^{66,67,68,70}$Zn continuously decreases toward higher metallicity.
The slope changes at [Fe/H] $\sim -2.5$ and $\sim -1.5$ are due to the onset of  intermediate- and 
low-mass AGB stars, respectively.
The rapid change in the slope at [Fe/H] $\sim -1$ is caused by SNe Ia.
In Table \ref{tabmodel}, we show the mass fractions of isotopes at [Fe/H] $=-2.6, -1.1$, and $0.0$, 
corresponding to the average of metal-poor SNe II, SNe II$+$AGB, and SNe II$+$AGB$+$SNe Ia, 
respectively, which can be compared with meteoric data in a future study.
The results are summarized as follows.

\begin{itemize}
\item {Carbon} ---
The $^{12}$C/$^{13}$C ratio is $\sim 4200$ at [Fe/H] $= -2.6$, and decreases to $109$ at [Fe/H] $= -1.1$ 
because of the production of $^{13}$C from $4-7 M_\odot$ AGB stars.
Then, because of the production of $^{12}$C from $1-4 M_\odot$ AGB stars, the ratio increases until 
[Fe/H] $\sim -0.5$, and then decreases to be $88.6$ at [Fe/H] $= 0$, which is consistent with the
solar ratio of $89.9$ (solar symbol, AG89) and $89.4$ \citep[hereafter AGSS09]{asp09}.
The isotopic fraction of $^{13}$C is 0.02\%, 0.84\%, 0.75\%, and 1.03\% at [Fe/H] $= -2.6, -1.1, -0.5$ and $0$, 
respectively.
The other dots (crosses and diamonds) show the observational data of metal-poor unevolved stars 
\citep{car00,spi06}. The low carbon isotopic values of these stars suggests that intermediate-mass AGB 
stars and/or rotating massive stars have contributed to galactic chemical evolution at a very 
low metallicity. Note that in the low-mass stellar models, non-standard {\em
extra mixing} processes are not included \citep[e.g. ][]{nollett03}. These processes may occur 
during the first and asymptotic giant branches and result in an increase the yields of
$^{13}$C (and $^{14}$N). While the effect of such processes may be minimal on the yields at solar 
metallicity, they may be substantial for $Z \ltsim 0.0001$, resulting in much smaller predicted
$^{12}$C/$^{13}$C ratios from the AGB models than given by the yields of \citet{kar10}. 

\item {Nitrogen} ---
Although $^{15}$N destruction by AGB stars is neglected in our models, the $^{14}$N/$^{15}$N ratio 
at [Fe/H] $=0$ is still predicted to be larger than the ratio provided by AG89 (272).
At [Fe/H] $=0$, the isotopic fraction of $^{15}$N is 0.04\%, which is $6-10$ times smaller than the 
solar ratio: 0.37\% in AG89 and 0.23\% in AGSS09. 
Note that the proto-solar nebula value for the $^{14}$N/$^{15}$N ratio in AGSS09 is 447, a factor
of 1.6 higher than the AG89 value. This is because AGSS09 adopt the Jupiter nitrogen isotope
value as the proto-solar value, noting that the AG89 ratio is the terrestrial value derived from air
and has likely experienced isotopic fractionation and an increase in the abundance of $^{15}$N 
\citep[see][for more details]{meibom07}.
That our predicted nitrogen isotope value is too high is probably because the effect of novae, 
which likely produced a substantial fraction of the $^{15}$N in the Galaxy \citep{rom03}, are not included in our models.
In novae, the accreted hydrogen is heated up to $\sim 2-3 \times 10^8$ K, where the CNO cycle is 
limited by $\beta$-decays rather than the proton capture rate of $^{14}$N (the hot CNO burning, \citealt{war81}). 
Therefore, $^{13}$C, $^{15}$N, and $^{17}$O are over-produced with respect to the solar abundances.
The nucleosynthesis yields depend on the mass of the CO and ONe white dwarfs, and the mixing levels 
between the accreted envelope and the white dwarfs \citep{jos98}. 
The rate of novae in the Galaxy is estimated to be about 30 per year \citep{jos98}, but the time evolution of the rate is uncertain.

\item {Oxygen} ---
Both the $^{16}$O/$^{17}$O and $^{16}$O/$^{18}$O ratios rapidly decrease from $\sim 190000$ and 
$\sim 14000$ at [Fe/H] $= -2.6$ to $1787$ and $457$ at [Fe/H] $= 0$ because of the metallicity 
dependence of massive star nucleosynthesis.
The isotopic fractions are ($^{16}$O:$^{17}$O:$^{18}$O) = (99.99, 0.0005, 0.006), (99.96, 0.006, 0.037), 
(99.87, 0.026, 0.109), and (99.75, 0.053, 0.194) at [Fe/H] $= -2.6, -1.1, -0.5$, and $0$, respectively.
At [Fe/H]=0, the $^{18}$O fraction is consistent but $^{17}$O is too large when compared with the solar 
ratios, (99.76, 0.038, 0.201) in AG89 and (99.76, 0.038, 0.200) in AGSS09.
$^{17}$O and $^{18}$O are mainly produced by
$^{16}{\rm O}(p,\gamma)^{17}{\rm F}(\beta^+)^{17}{\rm O}$ in the H-burning layer
 and
$^{14}{\rm N}(\alpha,\gamma)^{18}{\rm F}(\beta^+)^{18}{\rm O}$ in the He-burning layer, respectively,
and thus their abundances depend on the amount of the seed element.
$^{18}$O is mainly produced by He-burning in massive stars and is slightly destroyed in AGB stars.
$^{17}$O is over produced by AGB stars. 
The model that adopts only supernova yields (dashed line in Fig. \ref{fig:iso2}) produces a $^{16}$O/$^{17}$O ratio
that is consistent with the solar ratio.  Including the contribution from AGB stars lowers the 
predicted $^{16}$O/$^{17}$O ratio to a value lower than the solar ratio. 
Note that for the majority of AGB stars in galaxies ($Z\ge0.004$), the $^{17}$O yield is increased 
with the new reaction rates (\S 2). The oxygen isotopic ratios may be used to put constraints on 
the rates of the $^{17}$O(p,$\alpha$)$^{14}$N and $^{17}$O(p,$\gamma$)$^{18}$F reactions in AGB nucleosynthesis
models. Novae also produce some $^{17}$O (and $^{13}$C), which would worsen the situation.

\item {Neon} ---
The $^{20}$Ne/$^{21}$Ne and $^{20}$Ne/$^{22}$Ne ratios show a similar decrease as oxygen.
With AGB yields, the isotopic fractions at [Fe/H]= $0$ are consistent with the solar ratio:
($^{20}$Ne:$^{21}$Ne:$^{22}$Ne) = (96.16, 0.23, 3.61) in the model, 
(92.99, 0.23, 6.78) in AG89, and (92.94, 0.22, 6.83) in AGSS09.
The only data for comparison is meteoritic. 

\item {Magnesium} ---
The $^{24}$Mg/$^{25}$Mg and $^{24}$Mg/$^{26}$Mg ratios also show a rapid decrease from $247$ and 
$271$ at [Fe/H] $= -2.6$ to $8.54$ and $9.22$ at [Fe/H] $= 0$ because the production of the minor 
isotopes increases in metal-rich supernovae.
The $^{24}$Mg/$^{25}$Mg and $^{24}$Mg/$^{26}$Mg ratios are a bit larger than the observations of 
stars \citep[dots, ][]{yon03,mel07} and the solar ratios ($7.92$ and $7.19$ in AG89), which 
suggests that AGB stars (or Wolf-Rayet stars) need to contribute more at all metallicities.
The isotopic fractions are ($^{24}$Mg:$^{25}$Mg:$^{26}$Mg) = (99.28, 0.39, 0.34), (95.60, 2.30, 2.11), 
(88.09, 5.77, 5.13), and (82.46, 9.28, 8.26) at [Fe/H] $= -2.6, -1.1, -0.5$, and $0$, respectively.
The isotopic fractions of solar ratios are (79.03, 9.97, 10.99) in AG89 and (78.99, 10.00, 11.01) in AGSS09.
In both observations, $^{26}$Mg is more abundant than $^{25}$Mg, which is the opposite both for the AGB 
stars and supernovae yields.
In supernovae, there is a metallicity dependence on the predicted Mg isotopic ratios.
$^{24}$Mg is mainly produced by the reaction $^{12}$C$+^{12}$C and is a primary isotope. 
On the other hand, $^{25}$Mg and $^{26}$Mg are secondary isotopes and are significantly 
produced by higher metallicity stars. They are mainly produced by the following reactions: 
$^{24}{\rm Mg}(p,\gamma)^{25}{\rm Al}(\beta^+)^{25}{\rm Mg}(p,\gamma)^{26}{\rm Al}(\beta^+)^{26}{\rm Mg}$ 
and thus their abundances depend on the amount of the seed element, $^{24}$Mg.
In AGB stars, there is no such metallicity dependence.
The neutron-rich Mg isotopes are produced by $\alpha$-captures onto $^{22}$Ne via 
$^{22}$Ne($\alpha,n$)$^{25}$Mg and $^{22}$Ne($\alpha,\gamma$)$^{26}$Mg \citep{karakas03}.
The AGB yields of the neutron-rich Mg isotopes depends on the amount 
of $^{22}$Ne in the He-burning region, and this can have a primary 
component in intermediate-mass AGB stars, where the $^{22}$Ne is
produced from primary nitrogen made by HBB. 

\item {Si, S, Ca, Ti, and Zn} ---
For heavier elements most of the ratios at [Fe/H] $= 0$ are roughly consistent with the solar ratios, 
but the predictions for $^{29}$Si, $^{48}$Ca, $^{47}$Ti are smaller than the solar ratios.
These may require the update of the mixing treatment and reaction rates in the supernovae calculations.
For Zn, the offsets from the solar ratios are possibly caused by the under-production of $^{64}$Zn.
Although neutron-rich isotopes of Zn could be produced by neutron-capture processes, $^{64}$Zn is 
mostly produced by the higher energy and entropy experienced during supernova explosions.
In fact, if we were to set a higher fraction of hypernovae, the isotopic ratios become 
closer to the solar ratios.

\end{itemize}

The model dependence on the evolution of the isotope ratios is shown in Figure \ref{fig:iso2}.
With the contribution from AGB stars (solid lines), the $^{12}$C/$^{13}$C and $^{20}$Ne/$^{22}$Ne 
ratios become consistent with the solar ratio, and $^{24}$Mg/$^{25,26}$Mg becomes slightly closer 
to the observations. However, as mentioned above, the $^{16}$O/$^{17}$O ratio is consistent with the 
model without an AGB contribution (dashed lines).
This is different from \citet{tim95}'s results, where the $^{16}$O/$^{17}$O ratio was consistent 
with the solar ratio without AGB yields and $^{18}$O is over-produced by supernovae. 
This is due to our use of updated supernova yields.
For heavier elements, there is only a $\sim 0.01$ dex difference around [Fe/H] $\sim 0$ 
for the models with and without AGB yields.
For $^{24}$Mg/$^{25,26}$Mg, the contribution of AGB stars are larger 
in \citet{fen03} than in our models, which may be due to differences in 
the adopted IMF (\S 2.3).

If we include rotating massive stars (dotted lines), the $^{12}$C/$^{13}$C ratio becomes much 
closer to the observations of metal-poor stars as shown by \citet{chi08}, but $^{16}$O/$^{17}$O 
cannot be improved when comparing with the solar ratio. Note that the N and Mg yields of rotating 
massive stars are not included. The small difference among the models with and without rotating 
massive stars is caused by the difference in the IMF ($M_{\rm u,2}=120M_\odot$ is adopted in 
the case with rotating massive stars instead of $M_{\rm u,2}=50M_\odot$).

The dependence on the star formation history is shown in Figure \ref{fig:iso3}.
For $^{12}$C/$^{13}$C, the first decrease is due to $^{13}$C production from intermediate-mass 
AGB stars, the next increase is due to the $^{12}$C production from low-mass AGB stars, and the 
following second decrease is due to SNe Ia. 
These modulations appear at a higher metallicity in the bulge (long-dashed lines) and thick 
disk (dot-dashed lines) than in the solar neighbourhood (solid lines).
On the other hand, these evolutionary changes appear at lower metallicity in the halo 
(short-dashed lines).
In general, the ratios between the major and minor isotopes such as $^{24}$Mg/$^{25,26}$Mg 
are smaller in the bulge and thick disk, and are larger in the halo because of the metallicity 
effect of supernovae. However, the $^{16}$O/$^{17}$O ratio in the halo is low due to the production of 
$^{17}$O from low-mass AGB stars, as also seen in the high [(C, F)/Fe] abundances.
Therefore, the isotopic ratios can be used as a tool to pick out the stars that form in a 
system with a low chemical enrichment efficiency.
This may be possible in our halo, but more likely in small satellite galaxies that were accreted 
onto our Milky Way Galaxy. In Table \ref{tabmodel}, the mass fractions of isotopes at 
[Fe/H] $=-0.5$ are provided for the solar neighbourhood, halo, bulge, and thick disk models.

\begin{table}
\center
\caption{The nucleosynthesis yields of core-collapse supernovae updated from \citet{kob06} in the ejecta in $M_\odot$.}
\label{tabsn}
\begin{tabular}{lccc}
\hline
$Z$	&	0.004	&	0.02	&	0.02	\\
$M$	&	18	&	25	&	25	\\
$E$	&	1	&	1	&	10	\\
\hline
$M_{\rm final}$	&	17.60	&	17.72	&	17.25	\\
$M_{\rm cut}$	&	2.14	&	1.80	&	1.84	\\
p	&	7.95E+00	&	5.32E+00	&	5.32E+00	\\
d	&	6.50E-14	&	2.07E-16	&	2.24E-12	\\
$^3$He	&	1.81E-04	&	2.03E-04	&	2.03E-04	\\
$^4$He	&	5.06E+00	&	5.35E+00	&	5.36E+00	\\
$^6$Li	&	9.30E-17	&	9.09E-19	&	1.84E-16	\\
$^7$Li	&	4.90E-11	&	2.55E-13	&	2.49E-10	\\
$^9$Be	&	9.78E-19	&	3.10E-19	&	1.64E-19	\\
$^{10}$B	&	6.52E-15	&	4.10E-11	&	4.10E-11	\\
$^{11}$B	&	3.16E-12	&	1.70E-10	&	1.73E-10	\\
$^{12}$C	&	1.58E-01	&	1.53E-01	&	1.36E-01	\\
$^{13}$C	&	6.03E-04	&	2.64E-04	&	2.61E-04	\\
$^{14}$N	&	1.42E-02	&	5.61E-02	&	5.62E-02	\\
$^{15}$N	&	5.62E-05	&	6.85E-05	&	9.18E-05	\\
$^{16}$O	&	1.55E+00	&	3.01E+00	&	2.94E+00	\\
$^{17}$O	&	1.04E-04	&	9.86E-04	&	9.83E-04	\\
$^{18}$O	&	1.14E-04	&	3.95E-03	&	2.29E-03	\\
$^{19}$F	&	1.39E-05	&	2.43E-04	&	1.73E-04	\\
$^{20}$Ne	&	1.40E-01	&	1.00E+00	&	7.37E-01	\\
$^{21}$Ne	&	3.92E-04	&	2.50E-03	&	2.87E-03	\\
$^{22}$Ne	&	4.02E-03	&	2.17E-02	&	1.86E-02	\\
$^{23}$Na	&	4.34E-04	&	2.96E-02	&	2.03E-02	\\
$^{24}$Mg	&	1.12E-01	&	1.91E-01	&	2.02E-01	\\
$^{25}$Mg	&	2.80E-03	&	3.59E-02	&	2.88E-02	\\
$^{26}$Mg	&	1.40E-03	&	3.04E-02	&	2.56E-02	\\
$^{27}$Al	&	5.22E-03	&	3.20E-02	&	2.79E-02	\\
$^{28}$Si	&	2.16E-01	&	3.43E-01	&	2.45E-01	\\
$^{29}$Si	&	1.83E-03	&	6.64E-03	&	9.80E-03	\\
$^{30}$Si	&	2.14E-03	&	5.01E-03	&	1.41E-02	\\
$^{31}$P	&	4.82E-04	&	1.82E-03	&	2.92E-03	\\
$^{32}$S	&	1.18E-01	&	1.81E-01	&	9.60E-02	\\
$^{33}$S	&	2.12E-04	&	8.20E-04	&	9.77E-04	\\
$^{34}$S	&	8.31E-04	&	3.34E-03	&	9.63E-03	\\
$^{36}$S	&	1.23E-06	&	3.29E-05	&	3.74E-05	\\
$^{35}$Cl	&	5.87E-05	&	3.63E-04	&	3.79E-04	\\
$^{37}$Cl	&	2.52E-05	&	3.02E-04	&	2.16E-04	\\
$^{36}$Ar	&	2.19E-02	&	2.82E-02	&	1.34E-02	\\
$^{38}$Ar	&	3.48E-04	&	2.69E-03	&	3.07E-03	\\
$^{40}$Ar	&	1.54E-07	&	2.99E-06	&	7.39E-06	\\
$^{39}$K	&	4.17E-05	&	2.66E-04	&	1.77E-04	\\
$^{40}$K	&	1.49E-08	&	5.55E-07	&	4.25E-07	\\
$^{41}$K	&	5.27E-06	&	4.58E-05	&	2.04E-05	\\
$^{40}$Ca	&	2.00E-02	&	1.98E-02	&	9.85E-03	\\
$^{42}$Ca	&	9.76E-06	&	7.93E-05	&	7.72E-05	\\
$^{43}$Ca	&	3.80E-07	&	3.66E-06	&	3.66E-06	\\
$^{44}$Ca	&	1.06E-05	&	3.76E-05	&	1.01E-04	\\
$^{46}$Ca	&	7.79E-08	&	1.53E-06	&	2.73E-06	\\
$^{48}$Ca	&	4.01E-07	&	2.17E-06	&	2.82E-06	\\
$^{45}$Sc	&	4.38E-07	&	3.69E-06	&	4.13E-06	\\
$^{46}$Ti	&	4.13E-06	&	2.43E-05	&	3.22E-05	\\
$^{47}$Ti	&	7.58E-07	&	4.99E-06	&	7.68E-06	\\
$^{48}$Ti	&	2.49E-04	&	2.09E-04	&	1.74E-04	\\
$^{49}$Ti	&	1.23E-05	&	1.05E-05	&	1.01E-05	\\
$^{50}$Ti	&	1.04E-06	&	1.51E-05	&	1.32E-05	\\
$^{50}$V	&	1.86E-08	&	9.94E-08	&	2.07E-07	\\
$^{51}$V	&	1.57E-05	&	1.20E-05	&	1.80E-05	\\
$^{50}$Cr	&	5.15E-05	&	5.04E-05	&	8.15E-05	\\
$^{52}$Cr	&	3.89E-03	&	3.50E-03	&	1.26E-03	\\
$^{53}$Cr	&	2.40E-04	&	9.95E-05	&	1.06E-04	\\
$^{54}$Cr	&	2.75E-06	&	3.00E-05	&	2.70E-05	\\
\hline
\end{tabular}
\end{table}
\begin{table}
\center
\begin{tabular}{lccc}
\hline
$Z$	&	0.004	&	0.02	&	0.02	\\
$M$	&	18	&	25	&	25	\\
$E$	&	1	&	1	&	10	\\
\hline
$^{55}$Mn	&	7.21E-04	&	3.47E-04	&	4.27E-04	\\
$^{54}$Fe	&	5.04E-03	&	2.63E-03	&	4.14E-03	\\
$^{56}$Fe	&	7.32E-02	&	8.46E-02	&	1.15E-01	\\
$^{57}$Fe	&	6.27E-04	&	8.79E-04	&	2.97E-03	\\
$^{58}$Fe	&	7.52E-05	&	8.91E-04	&	8.41E-04	\\
$^{59}$Co	&	3.58E-05	&	4.01E-04	&	5.08E-04	\\
$^{58}$Ni	&	5.50E-04	&	7.74E-04	&	1.25E-03	\\
$^{60}$Ni	&	1.03E-04	&	9.14E-04	&	3.68E-03	\\
$^{61}$Ni	&	1.48E-05	&	2.05E-04	&	5.36E-04	\\
$^{62}$Ni	&	5.33E-05	&	5.77E-04	&	7.22E-04	\\
$^{64}$Ni	&	4.18E-05	&	8.40E-04	&	7.16E-04	\\
$^{63}$Cu	&	1.37E-05	&	3.65E-04	&	2.89E-04	\\
$^{65}$Cu	&	1.32E-05	&	2.20E-04	&	2.05E-04	\\
$^{64}$Zn	&	7.19E-06	&	5.65E-05	&	2.88E-04	\\
$^{66}$Zn	&	2.35E-05	&	3.34E-04	&	3.47E-04	\\
$^{67}$Zn	&	3.16E-06	&	8.10E-05	&	7.26E-05	\\
$^{68}$Zn	&	2.37E-05	&	4.63E-04	&	4.24E-04	\\
$^{70}$Zn	&	4.95E-07	&	1.14E-05	&	1.94E-05	\\
$^{69}$Ga	&	2.30E-06	&	5.77E-05	&	5.19E-05	\\
$^{71}$Ga	&	2.53E-06	&	3.36E-05	&	3.61E-05	\\
$^{70}$Ge	&	3.08E-06	&	7.28E-05	&	6.25E-05	\\
$^{72}$Ge	&	5.07E-06	&	1.19E-04	&	1.07E-04	\\
$^{73}$Ge	&	6.71E-07	&	2.73E-05	&	2.32E-05	\\
$^{74}$Ge	&	9.69E-06	&	3.34E-04	&	2.87E-04	\\
\hline
\end{tabular}
\end{table}

\section{Conclusions}

We have presented evolution of isotope ratios of elemental abundances (from C to Zn) in the solar neighbourhood,
bulge, halo, and thick disk, using chemical evolution models with the updated yields of AGB stars and 
core-collapse supernovae.  Although the ejected mass of heavy elements is much smaller for AGB stars than 
for supernovae, the AGB yields of C, N, and F are comparable to those of supernovae.
When examining the AGB yields, the isotopes of $^{12}$C and $^{19}$F are mainly produced in 
low-mass stars ($1-4 M_\odot$), whereas $^{13}$C, $^{14}$N, $^{22}$Ne, $^{25}$Mg, and $^{26}$Mg are mainly 
produced in intermediate-mass ($4-7 M_\odot$) stars.
By including AGB yields, the predicted ratios of [C/Fe], [N/Fe],
$^{12}$C/$^{13}$C, and $^{20}$Ne/$^{22}$Ne at [Fe/H] $=0$ are much
improved compared with  the observed ratios in the solar neighbourhood.
For supernovae, minor isotope production increases with increasing metallicity. 
Therefore, the [(C, F)/Fe] and isotope ratios can be used as a cosmic clock along 
with the [$\alpha$/Fe] ratio.

Because of the effect of the progenitor mass and metallicity, the evolutionary history of 
elements varies for environments with different star formation histories.
In the bulge and thick disk, the star formation timescale is shorter than in the solar neighbourhood, 
and [$\alpha$/Fe] is higher and [Mn/Fe] is lower because of a lower contribution from SNe Ia. 
In contrast, the [(Na, Al, P, Cl, K, Sc, Cu, Zn)/Fe] ratios are higher because of the effect 
of metallicity. In other words, from these elemental abundance ratios, it is possible to 
select metal-rich stars ([Fe/H] $\gtsim -1$) that formed in the early Universe as a result 
of rapid star formation.

In the halo, the chemical enrichment timescale is longer than in the solar neighbourhood, 
and [(C, F)/Fe] and $^{12}$C/$^{13}$C are higher because of a stronger contribution from low-mass AGB stars.
The [$\alpha$/Fe] and [Mn/Fe] relations are the same as in the solar neighbourhood, but 
[(Na, Al, P, Cl, K, Sc, Cu, Zn)/Fe] are lower and other isotope ratios such as $^{16}$O/$^{18}$O 
and $^{24}$Mg/$^{25,26}$Mg are larger because of the low metallicity.
Not only from elemental abundance ratios, but also from isotopic ratios, it is possible to 
select the stars that formed in a system with a low chemical enrichment efficiency.

Isotopic ratios provide useful information to improve theoretical predictions in terms of 
reaction rates, nucleosynthesis, and the modelling of convective mixing.
The under-production of $^{15}$N suggests the contribution from novae, the over-production 
of $^{17}$O requires the updating of reaction rates that feed into AGB yields, and 
the under-production of $^{29}$Si, $^{48}$Ca, $^{47}$Ti and possibly $^{64}$Zn may 
require the updating of supernova models. 
The observational constraints at low-metallicity are particularly important.
However, it is not easy to estimate isotopic ratios from observations of stellar spectra.
To detect the small line shifts between isotopes, very high quality data is required.
The next generation of large telescopes (e.g., the Giant Magellan Telescope and the European Extremely Large Telescope) 
will be constrained by the same problems but will be able to measure isotopic ratios 
out to larger distances. It will then be possible to study the evolution of isotopic ratios
in the local neighbourhood of our Milky Way Galaxy including in dwarf spheroidal galaxies.
These data will help to put important constraints on the star formation and chemical enrichment 
histories of such systems, and which may lead to the verification of the hierarchical 
clustering in CDM cosmology using chemodynamical simulations.

\section*{Acknowledgements}
We would like to thank K. Nomoto, D. Yong, M. Lugaro, and P. Gil-Pons for fruitful discussions.
This work was supported by the NCI National Facility at the Australian National University and 
the Institute for the Physics and Mathematics of the Universe, University of Tokyo.

\begin{table*}
\caption{The mass fractions of isotopes for the models of the solar neighbhorhood, halo, bulge, and thick disk.}
\label{tabmodel}
\begin{tabular}{lcccccccc}
\hline
Model	&	Solar	&	Solar	&	Solar	&	Solar	&	Halo	&	Bulge	&	Thick	 \\
${\rm [Fe/H]}$	&	-2.6	&	-1.1	&	-0.5	&	0	&	-0.5	&	-0.5	&	-0.5	 \\
\hline
p	&	7.60E-01	&	7.53E-01	&	7.44E-01	&	7.29E-01	&	7.34E-01	&	7.38E-01	&	7.41E-01	 \\
d	&	5.06E-05	&	4.96E-05	&	4.82E-05	&	4.61E-05	&	4.90E-05	&	4.72E-05	&	4.78E-05	 \\
$^3$He	&	2.28E-05	&	2.34E-05	&	2.88E-05	&	4.28E-05	&	9.10E-05	&	2.22E-05	&	2.67E-05	 \\
$^4$He	&	2.40E-01	&	2.43E-01	&	2.47E-01	&	2.53E-01	&	2.56E-01	&	2.49E-01	&	2.48E-01	 \\
$^6$Li	&	1.37E-13	&	1.34E-13	&	1.30E-13	&	1.25E-13	&	1.29E-13	&	1.28E-13	&	1.29E-13	 \\
$^7$Li	&	6.38E-10	&	6.54E-10	&	6.11E-10	&	5.83E-10	&	6.64E-10	&	6.09E-10	&	6.13E-10	 \\
$^9$Be	&	1.31E-17	&	1.30E-17	&	1.29E-17	&	1.29E-17	&	1.27E-17	&	1.29E-17	&	1.29E-17	 \\
$^{10}$B	&	6.64E-17	&	1.81E-14	&	1.27E-13	&	4.19E-13	&	7.36E-14	&	2.54E-13	&	1.59E-13	 \\
$^{11}$B	&	3.54E-16	&	2.29E-12	&	5.67E-12	&	5.96E-12	&	3.88E-12	&	6.68E-12	&	6.17E-12	 \\
$^{12}$C	&	5.75E-06	&	2.85E-04	&	8.55E-04	&	1.49E-03	&	2.02E-03	&	7.30E-04	&	9.42E-04	 \\
$^{13}$C	&	1.37E-09	&	2.61E-06	&	7.04E-06	&	1.68E-05	&	1.28E-05	&	7.63E-06	&	7.98E-06	 \\
$^{14}$N	&	1.08E-07	&	1.12E-04	&	3.20E-04	&	6.51E-04	&	3.69E-04	&	3.98E-04	&	3.65E-04	 \\
$^{15}$N	&	4.23E-11	&	2.15E-08	&	1.02E-07	&	2.80E-07	&	7.95E-08	&	1.86E-07	&	1.24E-07	 \\
$^{16}$O	&	7.93E-05	&	2.23E-03	&	5.18E-03	&	9.73E-03	&	4.45E-03	&	7.82E-03	&	6.09E-03	 \\
$^{17}$O	&	4.23E-10	&	1.41E-07	&	1.44E-06	&	5.44E-06	&	1.88E-06	&	2.31E-06	&	1.70E-06	 \\
$^{18}$O	&	5.52E-09	&	9.21E-07	&	6.36E-06	&	2.13E-05	&	3.81E-06	&	1.34E-05	&	8.07E-06	 \\
$^{19}$F	&	2.10E-11	&	4.34E-08	&	1.51E-07	&	2.99E-07	&	3.49E-07	&	1.19E-07	&	1.72E-07	 \\
$^{20}$Ne	&	1.25E-05	&	4.53E-04	&	1.21E-03	&	2.46E-03	&	9.34E-04	&	1.90E-03	&	1.43E-03	 \\
$^{21}$Ne	&	2.73E-09	&	3.53E-07	&	1.98E-06	&	6.18E-06	&	1.25E-06	&	3.93E-06	&	2.47E-06	 \\
$^{22}$Ne	&	1.09E-08	&	1.71E-05	&	5.39E-05	&	1.01E-04	&	1.08E-04	&	3.20E-05	&	6.01E-05	 \\
$^{23}$Na	&	5.23E-08	&	4.76E-06	&	2.37E-05	&	7.00E-05	&	1.59E-05	&	4.46E-05	&	2.92E-05	 \\
$^{24}$Mg	&	5.77E-06	&	1.68E-04	&	3.57E-04	&	6.32E-04	&	3.17E-04	&	5.23E-04	&	4.19E-04	 \\
$^{25}$Mg	&	2.34E-08	&	4.20E-06	&	2.41E-05	&	7.40E-05	&	1.54E-05	&	4.65E-05	&	2.99E-05	 \\
$^{26}$Mg	&	2.13E-08	&	4.01E-06	&	2.23E-05	&	6.86E-05	&	1.40E-05	&	4.36E-05	&	2.77E-05	 \\
$^{27}$Al	&	1.60E-07	&	8.45E-06	&	3.14E-05	&	8.39E-05	&	2.18E-05	&	5.65E-05	&	3.82E-05	 \\
$^{28}$Si	&	7.49E-06	&	2.07E-04	&	4.96E-04	&	9.67E-04	&	4.35E-04	&	6.82E-04	&	5.64E-04	 \\
$^{29}$Si	&	3.79E-08	&	2.07E-06	&	8.42E-06	&	2.37E-05	&	5.79E-06	&	1.52E-05	&	1.02E-05	 \\
$^{30}$Si	&	4.45E-08	&	2.48E-06	&	9.89E-06	&	2.78E-05	&	6.98E-06	&	1.74E-05	&	1.19E-05	 \\
$^{31}$P	&	1.55E-08	&	6.46E-07	&	2.53E-06	&	7.32E-06	&	1.83E-06	&	4.56E-06	&	3.07E-06	 \\
$^{32}$S	&	2.93E-06	&	8.61E-05	&	2.16E-04	&	4.43E-04	&	1.88E-04	&	2.95E-04	&	2.45E-04	 \\
$^{33}$S	&	8.42E-09	&	3.37E-07	&	1.10E-06	&	2.79E-06	&	8.70E-07	&	1.69E-06	&	1.27E-06	 \\
$^{34}$S	&	2.06E-08	&	1.43E-06	&	6.53E-06	&	1.91E-05	&	4.49E-06	&	1.14E-05	&	7.80E-06	 \\
$^{36}$S	&	6.34E-12	&	2.70E-09	&	2.70E-08	&	9.63E-08	&	1.48E-08	&	5.82E-08	&	3.43E-08	 \\
$^{35}$Cl	&	2.41E-09	&	9.94E-08	&	4.03E-07	&	1.18E-06	&	2.97E-07	&	6.95E-07	&	4.81E-07	 \\
$^{37}$Cl	&	8.51E-10	&	3.80E-08	&	2.08E-07	&	6.56E-07	&	1.35E-07	&	3.95E-07	&	2.54E-07	 \\
$^{36}$Ar	&	4.26E-07	&	1.33E-05	&	3.40E-05	&	7.03E-05	&	2.95E-05	&	4.57E-05	&	3.83E-05	 \\
$^{38}$Ar	&	8.92E-09	&	5.49E-07	&	2.56E-06	&	7.53E-06	&	1.80E-06	&	4.31E-06	&	3.01E-06	 \\
$^{40}$Ar	&	1.41E-12	&	5.20E-10	&	3.78E-09	&	1.25E-08	&	2.15E-09	&	7.75E-09	&	4.74E-09	 \\
$^{39}$K	&	1.20E-09	&	5.39E-08	&	2.10E-07	&	5.79E-07	&	1.56E-07	&	3.40E-07	&	2.45E-07	 \\
$^{40}$K	&	2.87E-13	&	2.85E-11	&	2.70E-10	&	9.81E-10	&	1.52E-10	&	5.87E-10	&	3.45E-10	 \\
$^{41}$K	&	2.08E-10	&	6.33E-09	&	2.63E-08	&	7.64E-08	&	1.90E-08	&	4.61E-08	&	3.15E-08	 \\
$^{40}$Ca	&	3.36E-07	&	1.07E-05	&	2.74E-05	&	5.57E-05	&	2.38E-05	&	3.63E-05	&	3.07E-05	 \\
$^{42}$Ca	&	2.18E-10	&	1.35E-08	&	6.52E-08	&	1.98E-07	&	4.70E-08	&	1.08E-07	&	7.60E-08	 \\
$^{43}$Ca	&	5.41E-12	&	6.54E-10	&	3.68E-09	&	1.15E-08	&	2.25E-09	&	7.20E-09	&	4.56E-09	 \\
$^{44}$Ca	&	1.51E-09	&	4.75E-08	&	1.42E-07	&	3.44E-07	&	1.09E-07	&	2.39E-07	&	1.70E-07	 \\
$^{46}$Ca	&	6.61E-13	&	1.45E-10	&	1.25E-09	&	4.38E-09	&	7.04E-10	&	2.67E-09	&	1.59E-09	 \\
$^{48}$Ca	&	2.02E-12	&	4.91E-10	&	3.54E-09	&	1.18E-08	&	2.04E-09	&	7.24E-09	&	4.45E-09	 \\
$^{45}$Sc	&	8.32E-12	&	5.67E-10	&	3.06E-09	&	9.54E-09	&	1.93E-09	&	5.93E-09	&	3.78E-09	 \\
$^{46}$Ti	&	1.96E-10	&	8.32E-09	&	3.18E-08	&	9.00E-08	&	2.48E-08	&	5.03E-08	&	3.68E-08	 \\
$^{47}$Ti	&	2.42E-10	&	5.89E-09	&	1.53E-08	&	3.64E-08	&	1.29E-08	&	2.58E-08	&	1.86E-08	 \\
$^{48}$Ti	&	4.40E-09	&	1.60E-07	&	4.54E-07	&	1.03E-06	&	3.79E-07	&	6.43E-07	&	5.16E-07	 \\
$^{49}$Ti	&	1.51E-10	&	6.47E-09	&	2.24E-08	&	5.87E-08	&	1.85E-08	&	3.09E-08	&	2.49E-08	 \\
$^{50}$Ti	&	6.55E-12	&	1.55E-09	&	1.36E-08	&	4.98E-08	&	9.71E-09	&	2.30E-08	&	1.57E-08	 \\
$^{50}$V	&	2.08E-13	&	2.26E-11	&	1.29E-10	&	4.15E-10	&	8.01E-11	&	2.56E-10	&	1.60E-10	 \\
$^{51}$V	&	5.35E-10	&	1.61E-08	&	5.12E-08	&	1.37E-07	&	4.68E-08	&	6.50E-08	&	5.57E-08	 \\
$^{50}$Cr	&	1.07E-09	&	4.00E-08	&	1.68E-07	&	5.23E-07	&	1.58E-07	&	1.86E-07	&	1.73E-07	 \\
$^{52}$Cr	&	5.05E-08	&	1.90E-06	&	5.66E-06	&	1.35E-05	&	5.06E-06	&	6.86E-06	&	6.10E-06	 \\
$^{53}$Cr	&	2.96E-09	&	1.18E-07	&	4.54E-07	&	1.36E-06	&	4.41E-07	&	4.51E-07	&	4.55E-07	 \\
$^{54}$Cr	&	1.78E-11	&	3.62E-09	&	6.18E-08	&	2.73E-07	&	6.43E-08	&	5.08E-08	&	5.63E-08	 \\
$^{55}$Mn	&	8.33E-09	&	3.54E-07	&	2.61E-06	&	1.04E-05	&	2.91E-06	&	1.59E-06	&	2.28E-06	 \\
\hline
\end{tabular}
\end{table*}
\begin{table*}
\begin{tabular}{lcccccccc}
\hline
Model	&	Solar	&	Solar	&	Solar	&	Solar	&	Halo	&	Bulge	&	Thick	 \\
${\rm [Fe/H]}$	&	-2.6	&	-1.1	&	-0.5	&	0	&	-0.5	&	-0.5	&	-0.5	 \\
\hline
$^{54}$Fe	&	7.24E-08	&	2.88E-06	&	2.81E-05	&	1.21E-04	&	3.33E-05	&	1.24E-05	&	2.31E-05	 \\
$^{56}$Fe	&	2.92E-06	&	9.63E-05	&	3.62E-04	&	1.11E-03	&	3.57E-04	&	3.79E-04	&	3.69E-04	 \\
$^{57}$Fe	&	4.58E-08	&	1.65E-06	&	9.97E-06	&	3.78E-05	&	1.05E-05	&	8.41E-06	&	9.41E-06	 \\
$^{58}$Fe	&	5.82E-10	&	1.35E-07	&	1.10E-06	&	3.75E-06	&	8.31E-07	&	1.62E-06	&	1.24E-06	 \\
$^{59}$Co	&	6.10E-09	&	2.03E-07	&	8.70E-07	&	2.66E-06	&	7.72E-07	&	1.11E-06	&	9.43E-07	 \\
$^{58}$Ni	&	2.10E-08	&	7.49E-07	&	2.80E-05	&	1.38E-04	&	3.58E-05	&	5.95E-06	&	2.08E-05	 \\
$^{60}$Ni	&	7.33E-08	&	2.25E-06	&	7.93E-06	&	2.33E-05	&	7.52E-06	&	9.44E-06	&	8.42E-06	 \\
$^{61}$Ni	&	1.38E-09	&	1.20E-07	&	5.77E-07	&	1.69E-06	&	4.80E-07	&	7.97E-07	&	6.56E-07	 \\
$^{62}$Ni	&	1.75E-09	&	1.31E-07	&	1.20E-06	&	4.79E-06	&	1.08E-06	&	1.45E-06	&	1.22E-06	 \\
$^{64}$Ni	&	3.00E-10	&	6.74E-08	&	4.93E-07	&	1.65E-06	&	2.82E-07	&	1.03E-06	&	6.21E-07	 \\
$^{63}$Cu	&	2.71E-10	&	2.53E-08	&	1.80E-07	&	6.05E-07	&	1.05E-07	&	3.75E-07	&	2.27E-07	 \\
$^{65}$Cu	&	1.22E-10	&	2.48E-08	&	1.52E-07	&	4.80E-07	&	8.94E-08	&	3.06E-07	&	1.90E-07	 \\
$^{64}$Zn	&	5.71E-09	&	1.57E-07	&	3.69E-07	&	7.08E-07	&	3.12E-07	&	5.57E-07	&	4.34E-07	 \\
$^{66}$Zn	&	2.49E-10	&	4.32E-08	&	2.65E-07	&	8.42E-07	&	1.59E-07	&	5.23E-07	&	3.28E-07	 \\
$^{67}$Zn	&	2.33E-11	&	6.23E-09	&	5.10E-08	&	1.75E-07	&	2.86E-08	&	1.08E-07	&	6.45E-08	 \\
$^{68}$Zn	&	1.53E-10	&	3.94E-08	&	3.11E-07	&	1.05E-06	&	1.75E-07	&	6.55E-07	&	3.93E-07	 \\
$^{70}$Zn	&	1.03E-12	&	3.72E-10	&	7.89E-09	&	3.14E-08	&	4.11E-09	&	1.84E-08	&	1.03E-08	 \\
$^{69}$Ga	&	1.90E-11	&	5.71E-09	&	3.84E-08	&	1.23E-07	&	2.20E-08	&	7.81E-08	&	4.80E-08	 \\
$^{71}$Ga	&	1.71E-11	&	4.43E-09	&	2.90E-08	&	9.30E-08	&	1.68E-08	&	5.90E-08	&	3.63E-08	 \\
$^{70}$Ge	&	3.31E-11	&	8.59E-09	&	5.63E-08	&	1.80E-07	&	3.25E-08	&	1.14E-07	&	7.04E-08	 \\
$^{72}$Ge	&	4.03E-11	&	1.20E-08	&	8.95E-08	&	2.98E-07	&	5.06E-08	&	1.86E-07	&	1.13E-07	 \\
$^{73}$Ge	&	4.43E-12	&	1.93E-09	&	1.90E-08	&	6.75E-08	&	1.04E-08	&	4.12E-08	&	2.42E-08	 \\
$^{74}$Ge	&	7.76E-11	&	3.09E-08	&	2.75E-07	&	9.54E-07	&	1.52E-07	&	5.88E-07	&	3.49E-07	 \\
\hline
\end{tabular}
\end{table*}

\label{lastpage}

\bsp

\end{document}